\documentclass[twocolumn,usenatbib]{mn2e}

\usepackage{natbib}
\usepackage{amsbsy}
\usepackage{amssymb}
\usepackage{amsmath}

\usepackage{graphicx}

\newcommand{\be}{\begin{equation}}
\newcommand{\ee}{\end{equation}}
\newcommand{\bear}{\begin{eqnarray}}
\newcommand{\eear}{\end{eqnarray}}
\newcommand{\n}{\hat{n}}
\newcommand{\s}{\hat{s}}
\newcommand{\lam}{\hat{\lambda}}
\newcommand{\rx}{{\rm x}}

\newcommand{\rn}{{\rm n}}
\newcommand{\re}{{\rm e}}
\newcommand{\rp}{{\rm p}}

\newcommand{\rnp}{{\rm np}}

\begin{document}

\title[Superfluid instabilities and neutron star precession]
{Do superfluid instabilities prevent neutron star precession?}

\author[Glampedakis, Andersson \& Jones]{K. Glampedakis$^1$, N. Andersson$^2$ \& D.I. Jones$^2$ \\
\\
$^1$ SISSA/International School for Advanced Studies and INFN, via Beirut 2-4, 34014 Trieste, Italy \\
$^2$ School of Mathematics, University of Southampton, Southampton SO17 1BJ, UK}

\maketitle

\begin{abstract}
We discuss short wavelength (inertial wave) instabilities present in the standard
two-fluid neutron star model when there is sufficient relative flow along the superfluid neutron vortex array. 
We demonstrate that these instabilities may be triggered in precessing neutron stars, since
the angular velocity vectors of the neutron and proton fluids are misaligned during precession.
The presence of such an instability would render the standard, solid body rotation, model
for free precession inconsistent. Our results suggest that the standard (Eulerian) slow precession
that results for weak drag between the vortices and 
the charged fluid (protons and electrons) is not seriously constrained by the existence of the
instability. In contrast, the fast  precession, which results when vortices are strongly
coupled to the charged component, is generally unstable. This implies that fast 
precession may not be realised in astrophysical systems.

\end{abstract}


\section{Introduction}
\label{sec:intro}

Over the last few years, the modelling of neutron star free precession
has attracted considerable attention from theorists.  This has been
caused in large part by the reported observations of free precession
in PSR B1828-11 by \citet{1828}. There exist several other
precession candidates, most notably PSR B1642-03 \citep{shabanov}, 
but none are as convincing as PSR B1828-11, whose pulse
variations extend over multiple precession periods and are visible in
multiple channels (pulse time-of-arrival and beam shape).

This precessional motion is of great interest in several regards.
Firstly, PSR B1828-11 is the only clear precession candidate out of the
several thousand known pulsars, which begs the questions why is
precession so rare, and what has set PSR B1828-11 into precession in
the first place?  Secondly, the free precession timescale as extracted
from the observations is approximately 500 days (modulo a possible
factor of two up or down, depending on which of the observed
harmonics in the frequency spectrum is the fundamental one).  This is
extremely uncomfortable from the theoretical point of view, as the
canonical model of a neutron star containing one or more superfluid
components would suggest a much shorter precession timescale.  
Instead, this
long period free precession period seems to be consistent with the
simple wobble of a star deformed away from sphericity by
strains in its crust or by a magnetic field, with superfluidity
playing no role whatsoever \citep{akgun}.  It is this second
question regarding the apparent lack of the effects of superfluidity
with which we will mainly be concerned in this paper.

The basic picture is as follows.  Both on theoretical grounds, and in
order to explain the glitch phenomena, neutron stars are believed to
contain superfluid components whose rotation is achieved by their
forming arrays of vortices pointing along the rotation axis.  In a
pioneering paper \citet{shaham77} showed that in the limit where these vortices
attach or `pin' themselves tightly to the radio-emitting crustal phase
(as is required to explain the glitches), the classical free
precession of the star is greatly modified by the gyroscopic effect of
the superfluid, whose angular momentum vector would be fixed in the
crust's frame.  The resulting free precession period is
then approximately $P/ (I_s/I_1 + \epsilon)$, where $P$ is the spin
period, $I_s$ and $I_1$ the moments of inertia of the superfluid and
crustal phases, and $\epsilon$ the dimensionless non-sphericity in the
crust's moment of inertia tensor.  Note that by `crust' we really mean
the crystalline crust itself and any other part of the star coupled to
it on sufficiently short timescale to participate in its precession,
which probably includes the core proton fluid.

One region where neutron vortices are expected to interact strongly with
the crust is the \emph{inner crust}, where neutron and crustal matter
coexist.  The interaction here is expected to be sufficiently strong
that the vortices pin tightly to the crustal lattice, resulting in a
very short free precession period, in conflict with the
PSR 1828-11 observations.  A possible resolution to this problem was
suggested by \citet{cutler}, who argued that forces created by
the precessional motion itself would un-pin the vortices, restoring
long period free precession, consistent with the observations.  More
problematically, neutron vortices in the \emph{core} of the neutron
star are also expected to interact with the charged component.
Specifically, the interior magnetic field is composed of a large
number of fluxtubes (a consequence of the protons being condensed in a
type-II superconductor), which can interact with the vortices
via a process known as \emph{mutual friction}.  As was shown in detail
by \citet{swc}, even if this coupling force
isn't strong enough to completely pin the vortices, the free
precession timescale in the coupled system is likely to be much
shorter than the observed precession timescale.  The significance of
this result was emphasised recently by \citet{link03}, who argued that
the observed slow precession necessitated a profound change in our
view of neutron star interiors, requiring either that one or other (or
both) of the posited superfluid/superconducting phases do not exist, or
else that there are no regions in the interior where they exist
simultaneously, perhaps because the protons form a type-I
superconductor.

It is this line of argument we address in this paper.  The results
outlined above, leading to the prediction of fast precession, have all
relied implicitly on treating the superfluid and non-superfluid phases
as rigid bodies, each with a well defined angular velocity vector.
However, in reality the phases are better described by the fluid Euler
equations, the ansatz of rigidity having been made mainly for reasons
of tractability.  In this paper, which expands on initial results 
discussed by \cite{preclett}, we demonstrate that the relative
motion between the two phases that rigid body free precession
generates can give rise to a superfluid two-stream instability.  The
instability itself occurs in the inertial modes of the rotating
fluids, and is related to the experimentally observed instability of
Helium~II explained long ago by Glaberson and collaborators \citep{glab}.  
It is also intimately linked to the neutron star
instability described very recently by \citet{sidery07}.  As we
will argue in detail, this instability renders the free precession
rigid rotation ansatz inconsistent and makes all subsequent
conclusions concerning the interior composition of neutron stars
unsafe.

The plan of this paper is as follows.  In Section 2 we describe the
basic formulation of the two-fluid coupled system. Section 3 
provides  the plane-wave analysis and the instability criteria for 
inertial modes in
various regimes.  In Section 4 we discuss the relevance of these results for 
freely precessing neutron stars. The detailed precession modes are analysed in 
Appendix~A drawing heavily on the previous work
of \citet{swc}.  Our conclusions can be found in Section 5.


\section{Formulation}
\label{sec:form}

When viewed as a multifluid system, a neutron star core can be
modelled (at the simplest level) as a mixture of superfluid neutrons
and a conglomerate of superconducting protons and normal electrons.
In this picture it is assumed that the charged components are
electromagnetically coupled on a timescale which is short compared to
the dynamics that is being investigated. The charged plasma then forms
a charge neutral fluid, which we will loosely refer to as the
``protons'' in the following.  When required we will use the constituent
index $\rx$, which can be either $\rn$ or $\rp$, to identify the different
components.  A key property of the system is that the neutron and
``proton'' fluids do not flow independently with respect to each
other. One of the key coupling mechanisms is the vortex mediated
mutual friction, which acts dissipatively to prevent relative flow and
also affects the nature of the waves in the system.  The mutual
friction force appears as the result of interactions between the
neutron vortices and the charged component.  In the standard form, it
represents scattering of electrons off of the magnetic field
associated with the vortex core \citep{als,sidery06}, but
there may also be contributions due to the interaction between
vortices and magnetic fluxtubes in the charged fluid etcetera.  This
latter effect may be important since the protons are expected to
condense into a type II superconductor \citep{baym} in which the
magnetic flux is carried by fluxtubes.

The smooth-averaged hydrodynamical equations for our system are
\citep{sidery06}, in a rotating frame with angular velocity $\Omega_0^i$, 
\be (\partial_ t + v^j_\rn \nabla_j ) v_i^\rn = \nabla_i
\psi_\rn + 2 \epsilon_{ijk} v^j_\rn \Omega^k_0 + f_i^{\rm mf}
\label{eulern}
\ee
\be
 (\partial_ t  + v^j_\rp \nabla_j ) v_i^\rp = \nabla_i \psi_\rp + 2 \epsilon_{ijk} v^j_\rp \Omega^k_0
- \frac{1}{x_\rp}f_i^{\rm mf} + \nu_{\rm ee} \nabla^2 v^i_\rp
\label{eulerp}
\ee 
In these equations $v^i_\rx $ are the constituent velocities
and $x_\rp = \rho_\rp/\rho_\rn $ is the density ratio. Since it is expected that the proton to neutron ratio in the outer
neutron star core where the above equations apply is about $0.05$, $x_\rp$ may also be taken to represent the
proton fraction $\rho_\rp/(\rho_\rn+\rho_\rp)$. We will not be making a distinction between these
quantities in the following. In the two Euler equations, the scalar functions $\psi_\rx = \tilde{\mu}_\rx + \Phi $
represent the sums of specific chemical potentials and
the gravitational potential.  Since our focus will be on inertial waves, we can assume that the
two fluids are incompressible.  This means that we also have the two
continuity equations \be \nabla_i v^i_\rx = 0
\label{cont}
\ee

In writing the above Euler equations we have made a number of
simplifying assumptions.  We have ignored the entrainment effect
between neutrons and protons\footnote{Even though the entrainment, due to the strong interaction 
between neutrons and protons, is an important effect we ignore it in order to simplify the analysis. 
Its inclusion would affect the quantitative results, but should not lead to 
any qualitative differences. It would, of course, be interesting to study 
the role of entrainment in this problem at some stage.} . We have assumed
that the charged components are comoving, which when combined with an
assumption of local charge neutrality, eliminates the electromagnetic
Lorentz force from the equations of motion.  We are also not
considering the effect of vortex tension or the elasticity of the vortex array, 
even though quantised
neutron vortices are present (in the volume averaged sense) in
(\ref{eulern}). The effect of the tension on the kinds of waves that
we will study has been considered by \cite{sidery07}, who found that
it was generally safe to ignore it. We are also not
accounting for the analogous fluxtube tension in (\ref{eulerp}).

We are, however, including the effect of shear viscosity in the
equations; this is important since we will consider very short wavelength dynamics. 
In a superfluid neutron star core, electron-electron
scattering provides the dominant contribution to the shear viscosity
\citep{flowers,viscous}.  This leads to the last term on the right-hand side of
(\ref{eulerp}). It should be noted that this description is expected
to be accurate at temperatures significantly below the various
superfluid transition temperatures. At higher temperatures one would
also have to account for various excitations, which will lead to the
presence of a number of additional viscous terms (see \citet{na_comer}
for a discussion).  In order to estimate the kinematic viscosity
coefficient in (\ref{eulerp}) we use the results of \citet{viscous}.
For a uniform density star with mass $1.4M_\odot$ and radius 10~km
(our canonical values) we then find that 
\be 
\nu_{\re\re} \approx 10^7
\left( {10^8~\mathrm{K} \over T} \right)^2\ \mathrm{cm}^2/\mathrm{s}
\label{viscous}
\ee

The mutual friction force in (\ref{eulern}) and (\ref{eulerp}) has the standard form
\citep{HV,sidery06}
\be
f_i^{\rm mf} =  {\cal B} \epsilon_{ijk} \epsilon^{kml} \hat{\omega}^j_\rn \omega_m^\rn w^\rnp_l
+  {\cal B}^\prime \epsilon_{ijk} \omega^j_\rn w^k_\rnp
\label{mf}
\ee
where $\omega^i_\rn = 2 \Omega_\rn^i+ \epsilon^{ijk} \nabla_j v^\rn_k$ is the neutron vorticity, the relative flow is
given by $w^i_\rnp = v^i_\rn -v^i_\rp$ and a `hat' denotes a unit vector.
This form for the mutual friction follows from assuming a resistive drag
force (per unit length)
\be
f^i_{\rm D} = \tilde{{\cal R}} ( v^i_{\rm p} - v^i_{\rm L})
\label{dragforce}
\ee
acting on each neutron vortex (moving with velocity $v^i_\mathrm{L}$).
Balancing the drag force against the standard Magnus force acting on the vortex leads to (\ref{mf}).
The dimensionless coupling coefficients ${\cal B}$ and ${\cal B}^\prime $
then follow from
\be
{\cal B} = \frac{{\cal R}}{1 + {\cal R}^2}, \qquad
\mbox{and} \qquad  {\cal B}^\prime = \frac{{\cal R}^2}{1 + {\cal R}^2}
\ee
where ${\cal R} = \tilde{{\cal R}} /\rho_\rn \kappa$. Here $\kappa=h/2m_\rn$ is the quantum of circulation.
This description is quite generic. It can be taken to account for
dissipative scattering of electrons off the magnetic field of the vortex core.
As discussed by, for example, \citet{sidery06} this leads to a
typical value of ${\cal B} \approx 4\times 10^{-4}$.
This means that we have
\be
{\cal B} \approx {\cal R} \ll 1  , \quad {\cal B}^\prime \approx {\cal B}^2 \ll {\cal B}
\label{weak}
\ee
We will refer to this as the {\it weak drag} limit.

However, it may well be the opposite limit that applies.
Strong coupling between the neutron and proton fluids may originate from the interaction of vortices
with the much more numerous magnetic fluxtubes \citep{sauls,ruderman,link03}. In this picture,
each vortex is ``pinned'' to a large number of fluxtubes. As a consequence it tends to  be forced to
move with the proton fluid. Essentially, this additional interaction can be accounted for
by modifying the drag force to
\be
f^i_{\rm D} = \tilde{{\cal R}} ( v^i_{\rm p} - v^i_{\rm L}) + \tilde{\cal R}' ( v_\phi^i - v^i_{\rm L})
\ee
where $ v_\phi^i$ is the velocity of the fluxtubes. If we assume that the fluxtubes move with the
electron/proton fluid, then it is clear that we will again end up with a mutual
friction force of the  form (\ref{mf}). Of course, this will not be the case if there is a slippage between
the fluxtubes and the charged component\footnote{Neither will (\ref{mf}) result if the drag coefficient $\tilde{\cal R}$ is velocity dependent.}. 
We will not consider this possibility here.
The precise form of the drag force experienced by the vortices in this case is not well known, but it is usually assumed 
that the pinning to the fluxtubes is efficient.  This would mean that one would have 
$\tilde{\cal R}' \gg \tilde{\cal R}$ and the final mutual friction force takes the form (\ref{mf}) with  
${\cal R} \approx  \tilde{\cal R}'/\rho_\rn \kappa \gg 1 $. In this case we have the {\it strong drag} limit,
\be
{\cal B} \approx \frac{1}{{\cal R}} \ll 1 \ , \quad   1-{\cal B}^\prime \approx {\cal B}^2 \ll 1
\ee

It is worth noting that some constraints on the value of ${\cal B} $ can be placed
by post-glitch relaxation data, using the two-fluid formalism 
(with the crust assumed rigidly attached to the ``protons''). For example, Vela's
1988 ``Christmas'' glitch implies a crust-core coupling timescale $\tau_{\rm c} \leq
300 (I_1/I_s)~\mbox{s} $ \citep{abney}. A simple glitch relaxation
model can be constructed from eqns.~(\ref{eulern})--(\ref{eulerp}), assuming rigid-body
rotation for the fluids. The resulting coupling timescale due to mutual friction
is found to be $\tau_{\rm mf} = (I_1/I_s)/(2\Omega {\cal B})$, in agreement with the
result of \citet{sidery06}. For a rotation period $P= 2\pi/\Omega \sim 0.1~\mbox{s}$
and $I_1 \approx 0.1 I_s $, the requirement $\tau_{\rm mf} < \tau_{\rm c} $ provides the constraint 
${\cal B} > 2 \times 10^{-5} $. This translates to either ${\cal R} >  2\times 10^{-5}$ 
(weak drag) or ${\cal R} < 4 \times 10^4 $ (strong drag). 

Finally, it is worth making a few remarks on the precession model that we
consider. Following \citet{swc} we assume that the global motion follows from (\ref{eulern}) and 
(\ref{eulerp}) after imposing that each fluid rotates as a solid body. Misaligning the two rotation vectors
we then arrive at a convenient two-component model for free precession which accounts for the damping 
due to mutual friction. The detailed precession modes are discussed in Appendix~A. This model represents
the current state-of-the-art in this problem area. However, it is obviously simplistic. The true
``layering'' of a real neutron star, and possible effects due to the various interfaces/boundaries
are not accounted for. This means that one must be careful when trying to connect 
our model to  
detailed calculations for the various parameters, e.g.  the mutual friction coefficients discussed by \citet{sidery06}. 
Basically, the model should  be seen as a {\em body averaged} 
representation of the real system, and the same is true for (say) the drag coefficient $\mathcal{R}$. 

\section{Local wave analysis}
\label{sec:local}

The present investigation is motivated by the fact that a relative
flow along a superfluid vortex array may trigger an instability that
leads to the formation of vortex tangles and a state of superfluid turbulence,
see \citet{turbulent,sidery07} for discussion and references to the
relevant literature.  Our aim is to find out whether this vortex
instability may be active in a precessing neutron star.

As a first step we consider perturbations with respect to a stationary background configuration where both fluids rotate
rigidly, i.e
\be
v^i_{\rx 0} = \epsilon^{ijk} (\Omega^\rx_j -\Omega_{0 j} )  x_k, \quad \Omega^i_\rx = \mbox{constant}
\ee
allowing for an arbitrary orientation of the angular velocity vectors.
After linearising and fixing $\Omega^i_0=\Omega^i_\rn = \Omega_\rn \n^i $, i.e. working in the neutron frame
(for later convenience), the various quantities appearing in the Euler equations
(\ref{eulern})-(\ref{eulerp}) become
\begin{displaymath}
v^i_\rx \to v^i_{\rx 0} +  v^i_\rx, \quad v^i_{\rn 0} = 0, \quad
v^i_{\rp 0}  = \epsilon^{ijk} (\Omega^\rp_j -\Omega^\rn_j) x_k
\end{displaymath}
\begin{displaymath}
 \omega^i_\rn = 2 \Omega_\rn^i + \epsilon^{ijk} \nabla_j v_k^\rn, \quad
\hat{\omega}^i_\rn = \n^i + \frac{1}{2\Omega_\rn} \left (\, g^{ij} -\n^i \n^j  \, \right ) \delta \omega_j^\rn
\end{displaymath}
\begin{displaymath}\\
\psi_\rx = \psi_{\rx 0} + \delta \psi_\rx
\end{displaymath}
We now focus our attention on short wavelength motion, such that we can analyse the dynamics without
considering boundary conditions etcetera. In practice, we use the standard
plane-wave decomposition
\be
v^i_\rx = A^i_\rx \, e^{i \sigma t + i k_j x^j}, \qquad A^i_\rx = \mbox{constant}
\label{plane}
\ee
in (\ref{eulern}), (\ref{eulerp}) and (\ref{cont}).
It is useful to decompose various vectors into a piece parallel to the vortex
array and a piece that is orthogonal to it, i.e. use
\be
A_\rx^i = A_\rx^\parallel\ \n^i + A_\rx^{\perp, i}
\ee
and similarly for all other variables. Given that we have assumed that the
fluids are incompressible, we must have
\be
k_j A_\rx^j = 0
\ee
In other words, the waves we consider are transverse.
To simplify the analysis somewhat, we will assume that the waves propagate
along the vortex array in the following. This means that we align
$k^i$ with $\hat{n}^i$, see \citet{sidery07} for comments on this assumption.
Hence we have
\be
k^i_\perp = 0 \quad \Rightarrow \quad k^i = k_\parallel \n^i
\ee
and
\be
\hat{n}_j A_\rx^j = 0 \quad \Rightarrow \quad A_\rx^\parallel = 0
\ee
Finally, we define two additional unit vectors $\s^i$ and $\lam^i $  to form an orthonormal triad,
\be
 \s^i \n_i =0 \qquad \mbox{and} \qquad   \lam^i = \epsilon^{ijk} \s_j \n_k
\ee

One of the key reasons why the above assumptions are useful is that we do not need to
worry about the perturbations in the potentials $\psi_\rx$. They are simply obtained
by projecting the Euler equations onto $\n^i$. To determine a dispersion relation for the
propagation and analyse the nature of the associated waves, it is sufficient to
consider the equations obtained by projecting the Euler equations in the directions
 $\s^i$ and $\lam^i$.

Finally, in order to write down the linearised version of the Euler equations (\ref{eulern})
and (\ref{eulerp}) we need an expression for the perturbed mutual friction force.
This can be written
\begin{multline}
\delta f^i_{\rm mf} = -2\Omega_\rn {\cal B} ( A^i_\rn - A^i_\rp) \\
+ i k_\parallel {\cal B}
\left ( \n^i \epsilon_{jkl} w_0^j \n^k A^l_\rn +
w_\parallel \epsilon^{ijk} \n_j A_k^\rn  \right )  \\
+ 2\Omega_\rn {\cal B}^\prime \epsilon^{ijk}
 \n_j (v^\rn_k -v^\rp_k) + i k_\parallel {\cal B}^\prime w_{0 j} ( \n^j A^i_\rn -\n^i A^j_\rn)
\end{multline}

\subsection{Restricted problem: perturbations in neutron fluid only}
\label{pinfo}

Before considering the general solutions to the above equations it is
useful to note that a simplified version of the present problem has
already been discussed by \citet{sidery07} (see \cite{glab} for the
analysis of the superfluid Helium case). They assume that the proton
fluid is locked to its background value, $v^i_{\rp 0} = \epsilon^{ijk}
(\Omega^\rp_j -\Omega^\rn_j) x_k $ in our case, and that it is
sufficient to consider only the neutron equation (\ref{eulern}).  It
is useful to recall the results obtained for this simplified problem
for two reasons.  First of all, the discussion explains our general
strategy.  Secondly, and more importantly, we will later demonstrate
that the results remain essentially unchanged in the weak drag limit.

Assuming that the protons remain unperturbed, we easily obtain
\begin{multline}
\left (\, i \sigma  + 2 \Omega_\rn {\cal B}  -i {\cal B}^\prime k_\parallel w_\parallel
\, \right ) A^i_\rn \\
+  \left [\, (1-{\cal B}^\prime) 2 \Omega_\rn
-i {\cal B} k_\parallel w_\parallel \, \right ]\epsilon^{ijk} \n_j A^\rn_k = 0
\label{1fluid}
\end{multline}
where
\be
w_\parallel = \n_i ( v^i_{\rn 0} - v^i_{\rp 0} ) = -\n_i v^i_{\rp 0}
\ee
Taking the inner product of
(\ref{1fluid}) with $v^i_\rn  $ gives
\be
(A_\perp^\rn)^2 = 0  \quad \Rightarrow \quad  \s_i A_{\rn}^i = \pm i ( \lam_i  A^i_{\rn} )
\ee
Hence the modes under discussion represent circularly polarised waves.
Next, projecting (\ref{1fluid}) along $\s^i$ and $\lam^i $ we get
\begin{multline}
\left [\, 2 {\cal B} \Omega_\rn + i \left (\, \sigma -{\cal B}^\prime k_\parallel w_\parallel \, \right ) \right ]
(\s_i A^i_\rn) \\
+ \left [\, 2 (1-{\cal B}^\prime) \Omega_\rn  -i{\cal B} w_\parallel k_\parallel \, \right ] (\lam_i A^i_\rn) = 0
\label{equ12a}
\end{multline}
\begin{multline}
\left [\, 2 (1-{\cal B}^\prime) \Omega_\rn -i{\cal B} w_\parallel k_\parallel \, \right ] (\s_i A^i_\rn) \\
- \left [\, 2{ \cal B} \Omega_\rn  + i \left (\, \sigma -{\cal B}^\prime k_\parallel w_\parallel \, \right ) \right ]
(\lam_i A^i_\rn)  = 0
\label{equ12b}
\end{multline}
Requiring a non-trivial solution for this system we find the dispersion relation,
\be
\sigma = \left [ \, {\cal B}^\prime k_\parallel w_\parallel \pm 2 (1-{\cal B}^\prime) \Omega_\rn \, \right ]
+ i {\cal B} \left( 2 \Omega_\rn \mp k_\parallel w_\parallel \, \right)
\label{disp1}
\ee
Note that this result is valid for arbitrary ${\cal B}$ and ${\cal B}'$, i.e. it is relevant
for
both the
strong and weak drag limits. It is also in agreement with the result found by
\citet{glab} in the limit of purely transverse waves, and \citet{sidery07} in the
limit ${\cal B}' \to 0$.

In the limit of vanishing mutual friction the dispersion relation reduces
to  $\sigma = \pm 2 \Omega_\rn $. These modes represent collective vortex Kelvin oscillations.
At the macroscopic level these waves can be identified with inertial waves in the neutron fluid \citep{sonin,sidery07}.
In the presence of mutual friction the inertial modes are generally damped but, remarkably,
they may become {\it unstable} when there is a relative flow along the vortex array.
We have an instability when
\be
\mbox{Im}\ \sigma  < 0 \quad \Rightarrow \quad  | w_\parallel | > \frac{2 \Omega_\rn}{k_\parallel}
\label{crit}
\ee
In the case of superfluid Helium this instability was first discussed by
\citet{glab}, and is often refereed to as the Donnelly-Glaberson instability \citep{turbook}.
In the context of neutron stars the instability was recently discussed by \citet{sidery07}, who
 argued that is a typical ``two-stream'' instability,
a generic property of multifluid systems (see \citet{twostream} for references to the relevant literature).
The relevance of this instability for neutron stars has also been discussed by \citet{peralta05,peralta06}. 

We want to consider these plane-wave results in the context of a
neutron star that is freely precessing. This means that we need to
connect the background that we have used in the plane-wave analysis to
the global motion of a precessing star. In all current models of
precession (see the discussion in Appendix~\ref{app:precess}) it is
assumed that the two components rotate as solid bodies. However, since
the rotation axes are not aligned one would in general expect both
$\Omega_\rn^i$ and $\Omega_\rp^i$ to be time dependent.  Since they are
taken to be constant in the plane-wave calculation it is obvious that
the result is only valid on a timescale that is short compared to the
variation in $\Omega_\rn^i$ and $\Omega_\rp^i$. As we will argue later,
the relevant timescale to compare to is the precession period. Hence
our analysis is applicable only if the instability growth time is
significantly shorter than the precession period. 

It is also clear that the unstable waves need to have relatively short wavelength.
This becomes obvious if we consider the explicit form for $w_\parallel $.
For our problem we have
\begin{multline}
w_\parallel = -\n_i v^i_{\rp 0} = - \n^i \epsilon_{ijk} \Omega^j_\rp x^k \\
 =  (\lam^i \Omega^\rp_i) (\s_j x^j) - (\s^i \Omega^\rp_i) (\lam_j x^j)
\end{multline}
This shows that the projection, $w_\parallel $, of the background flow along
the direction of the neutron vortices varies with position. This obviously does not invalidate
the plane-wave analysis, but it is clear that the characteristic
lengthscale must be much smaller than the radius of the star.

Finally, we need to appreciate that there will also be a cut-off at short wavelengths.
If we had accounted for the vortex tension in our analysis we
would have found a slightly different instability criterion \citep{glab,sidery07}
\be
 | w_\parallel | > \frac{2 \Omega_\rn}{k_\parallel} + \nu k_\parallel
\label{tenterm}\ee
where $\nu$ represents the tension. From this it is clear that the tension
tends to stabilise the system for large $k_\parallel$ (short wavelengths).
However, if we compare the relative magnitude of the
terms in (\ref{tenterm}) we immediately see that the tension
is rather insignificant. The two contributions are equal for a wavelength,
\begin{multline}
\nu k_\parallel^2 =  2\Omega_\rn \quad \Rightarrow \\
\lambda_0 = 2\pi \left ( \frac{\nu}{2\Omega_\rn} \right )^{1/2}  \approx 7 \times 10^{-2}\sqrt{P(s)} ~\mbox{cm}
\end{multline}
where $P(s)$ is the rotation period in seconds.
In other words, the tension contribution dominates only for wavelengths $\lambda_0$ comparable to the
intervortex spacing 
\be
d_\rn \approx 10^{-2} \sqrt{P(s)} ~\mbox{cm}
\ee
At that level the macroscopic hydrodynamics that we have been using is no longer relevant.
Equations (\ref{eulern})--(\ref{cont}) arrive after averaging over
lengthscales much larger than $d_\rn $. Consequently, the plane wave analysis makes sense provided that
$\lambda \gg d_\rn $, which also means that $\lambda \gg \lambda_0 $. Hence, as far as the present 
hydrodynamics problem is concerned, we can always neglect the vortex tension.
The vortex tension would be important if we were to consider the dynamics of a single vortex. 
In fact, an analysis of that problem shows that an oscillating vortex can also become unstable. 
The relevant instability criterion is similar to (\ref{tenterm}), although as expected for large $k_\parallel$ the first 
term is irrelevant and  
the tension plays the key role.


\subsection{Complete problem: perturbations in both fluids}

In a neutron star core there is no compelling argument for clamping
the proton fluid as we did in the previous example. It was a useful
assumption because it simplified the algebra. It also provided a
straightforward demonstration of the presence of unstable waves in the
system.  However, if we want to make contact with realistic precessing
neutron star models we need a more detailed analysis. Hence, we
consider the case where both fluids are perturbed.

In the general case we have the two equations of motion
\begin{multline}
\left ( i\sigma + 2 \Omega_\rn {\cal B} -i{\cal B}^\prime w_\parallel k_\parallel \right ) A^i_\rn
\\
-2\Omega_n {\cal B} A^i_\rp 
+ 2 \Omega_\rn {\cal B}^\prime \epsilon^{ijk} \n_j A_k^\rp \\
+ \left [ 2\Omega_\rn(1-{\cal B}^\prime) - i{\cal B} w_\parallel k_\parallel \right ] \epsilon^{ijk} \n_j A^\rn_k
=0
\label{equ_n}
\end{multline}
and
\begin{multline}
 \left [ i\sigma + v_{\rm ee} k^2_\parallel -i k_\parallel w_\parallel
+ 2\Omega_\rn \left (\frac{{\cal B}}{x_\rp}\right )   \right ] A^i_\rp \\
+
\left [ ik_\parallel w_\parallel  \left (\frac{{\cal B}^\prime}{x_\rp}\right )
- 2\Omega_\rn \left (\frac{{\cal B}}{x_\rp}\right )  \right ] A^i_\rn \\
+ \left [ ik_\parallel w_\parallel  \left (\frac{{\cal B}}{x_\rp}\right ) 
+ 2\Omega_\rn \left (\frac{{\cal B}^\prime}{x_\rp}\right )  \right ] \epsilon^{ijk} \n_j A_k^\rn
 \\
 + \left [ \Omega_\rn \left ( 1 -\frac{2{\cal B}^\prime}{x_\rp} \right ) + \Omega_\rp^\parallel  \right ]
\epsilon^{ijk} \n_j A^\rp_k  =0
\label{equ_p}
\end{multline}
where $\Omega_\rp^\parallel = \n_i \Omega^i_\rp $.
Projection of these equations along the $\s^i$ and $\lam^i $ vectors results in the $4\times4$
system
\begin{multline}
\left ( i\sigma + 2 {\cal B} \Omega_\rn -i{\cal B}^\prime w_\parallel k_\parallel  \right ) (\s_i A^i_\rn)
-2\Omega_\rn {\cal B} (\s_i A^i_\rp) \\
+ \left [ 2\Omega_\rn (1-{\cal B}^\prime) -i{\cal B} w_\parallel k_\parallel
\right ] (\lam_i A^i_\rn) + 2\Omega_\rn {\cal B}^\prime (\lam_i A^i_\rp) = 0
\label{equ1}
\end{multline}
\begin{multline}
\left ( i\sigma + 2 {\cal B}\Omega_\rn -i{\cal B}^\prime w_\parallel k_\parallel  \right ) (\lam_i A^i_\rn)
-2\Omega_\rn {\cal B} (\lam_i A^i_\rp) \\
- \left [ 2\Omega_\rn (1-{\cal B}^\prime) -i{\cal B} w_\parallel k_\parallel
\right ] (\s_i A^i_\rn) - 2\Omega_\rn {\cal B}^\prime (\s_i A^i_\rp) = 0
\end{multline}
\begin{multline}
\left [ i\sigma + \nu_{\rm ee} k^2_\parallel + 2\Omega_\rn \left ( \frac{{\cal B}}{x_\rp} \right )
-i k_\parallel w_\parallel \right ] (\s_i A^i_\rp) \\
+ \left [ ik_\parallel w_\parallel
\left ( \frac{{\cal B}^\prime}{x_\rp} \right ) - 2\Omega_\rn \left ( \frac{{\cal B}}{x_\rp} \right ) \right ]
(\s_i A^i_\rn)
 \\
 + \left [ 2\Omega_\rn \left ( \frac{{\cal B}^\prime}{x_\rp} \right ) + ik_\parallel w_\parallel
\left ( \frac{{\cal B}}{x_\rp} \right ) \right ] (\lam_i A^i_\rn) \\
+ \left [ \Omega_n \left ( 1 -2
\frac{{\cal B}^\prime}{x_\rp} \right ) + \Omega_\rp^\parallel \right ] (\lam_i A^i_\rp) = 0
\end{multline}
\begin{multline} \left [ i\sigma + \nu_{\rm ee} k^2_\parallel + 2\Omega_\rn \left ( \frac{{\cal B}}{x_\rp} \right )
-i k_\parallel w_\parallel \right ] (\lam_i A^i_\rp) \\
+ \left [ ik_\parallel w_\parallel
\left ( \frac{{\cal B}^\prime}{x_\rp} \right ) - 2\Omega_\rn \left ( \frac{{\cal B}}{x_\rp} \right ) \right ]
(\lam_i A^i_\rn)
 \\
 - \left [ 2\Omega_\rn \left ( \frac{{\cal B}^\prime}{x_\rp} \right ) + ik_\parallel w_\parallel
\left ( \frac{{\cal B}}{x_\rp} \right ) \right ] (\s_i A^i_\rn) \\
- \left [ \Omega_n \left ( 1 -2
\frac{{\cal B}^\prime}{x_\rp} \right ) + \Omega_\rp^\parallel \right ] (\s_i A^i_\rp) = 0
\label{equ4}
\end{multline}
The dispersion relation associated with this system is a fourth order polynomial in $\sigma$.
The roots can be explicitly calculated, but the dependence on the various parameters will be complicated.
In order to explore the nature of the waves in the system it makes sense to focus on
specific limiting cases. This way we also obtain useful approximate expressions
for the growth time of the unstable modes. As a first approximation it is natural to consider
the {\it inviscid} problem, i.e. we set $\nu_{\rm ee} = 0 $.


\subsubsection{Approximate inviscid solutions}

With the shear viscosity eliminated, we can arrive at different approximate solutions by
considering, separately, (i) the short wavelength limit $k_\parallel R \to \infty $ and
(ii) the weak/strong drag limits, ${\cal R} \to 0 $ and ${\cal R} \to \infty $,
respectively.

The short wavelength approximation leads to the following
mode solutions, valid for every ${\cal R} $,
\bear
&& \sigma_{1,2} \approx  \pm 2\Omega_\rn + ( i{\cal B} \mp {\cal B}^\prime ) \left ( 2\Omega_\rn
\mp k_\parallel w_\parallel \right )
\label{inv1}
\\
\nonumber \\
&& \sigma_{3,4} \approx \pm (\Omega_\rn + \Omega_\rp^\parallel) + k_\parallel w_\parallel
+\frac{2 \Omega_\rn}{x_\rp} (  i{\cal B}     \mp {\cal B}^\prime )
\label{inv2}
\eear
The top pair is identical to the modes found in the problem where the protons were kept clamped. These modes can still 
become unstable under the action of mutual friction, and the instability
criterion (\ref{crit}) remains {\it unchanged}.

For a generic value of  ${\cal R}$ both mode pairs in (\ref{inv1})-(\ref{inv2}) describe waves
where the neutron and proton fluids oscillate in concert. However, a closer look reveals their
distinct nature. In the weak drag limit, ${\cal R} \to 0 $, equations (\ref{equ_n}) and (\ref{equ_p}) decouple, and
the modes (\ref{inv1}) represent  {\it pure} neutron fluid inertial waves, circularly polarised.
The second pair of modes (\ref{inv2}) are different, representing the proton inertial modes as viewed in a frame rotating
with the neutrons. These modes are always damped by mutual friction.

Next, we derive approximate solutions for weak and strong drag without any limitation on the wavelength.
The ${\cal R} \to 0 $ case is the easiest. Retaining only the leading order mutual friction, the modes have the form
(\ref{inv1})-(\ref{inv2}) with ${\cal B}^\prime =0 $. As we have just discussed, these are neutron
inertial waves weakly coupled to the protons and vice-versa.

The strong coupling solutions are obtained after letting
${\cal B}^\prime \to 1$ and ${\cal B} \to 0 $ in the dispersion relation. Keeping the
same identification for the modes as in (\ref{inv1})-(\ref{inv2}), we find
\begin{multline}
\sigma_{1,2} \approx \pm { 1 \over 2} \left[ \Omega_\rp^\parallel + \left( 1-{2 \over x_\rp} \right) \Omega_\rn \right]
- k_\parallel w_\parallel \\
- { 1 \over 2 x_\rp} \Big\{ (\Omega_\rp^\parallel+\Omega_\rn)^2 x_\rp^2 
+4 \Omega_\rn[ (1+ 3x_\rp) \Omega_\rn - x_\rp \Omega_\rp^\parallel] \\
\mp 8 x_\rp \Omega_\rn k_\parallel w_\parallel
\Big\}^{1/2}
\end{multline}
\begin{multline}
\sigma_{3,4} \approx \pm { 1 \over 2} \left[ \Omega_\rp^\parallel + \left( 1-{2 \over x_\rp} \right) \Omega_\rn \right]
- k_\parallel w_\parallel \\
+ { 1 \over 2 x_\rp} \Big\{ (\Omega_\rp^\parallel+\Omega_\rn)^2 x_\rp^2
+4 \Omega_\rn[ (1+ 3x_\rp) \Omega_\rn - x_\rp \Omega_\rp^\parallel] \\
\mp 8 x_\rp \Omega_\rn k_\parallel w_\parallel
\Big\}^{1/2}
\end{multline}
It is clear that the strong drag coupling alters the waves in the system significantly.
It is no longer the case that the neutrons and protons can oscillate ``independently''.
Moreover, the above results show that the system may now be unstable already at leading order 
(i.e. when ${\cal B}^\prime = 1$ and ${\cal B} = 0$).
In order to discuss this strong drag instability further we assume that
 $\Omega^i_\rp$ and $\Omega^i_\rn$ are almost aligned. As we will see later, this assumption is valid for the
free precession motion. Setting
$\Omega_\rp^\parallel \approx \Omega_\rn $ the mode-frequencies become
\begin{multline}
\sigma_{1,2} \approx \pm \Omega_\rn \left ( 1 -\frac{1}{x_\rp} \right ) - k_\parallel w_\parallel
\\
- \frac{1}{x_\rp} \left [ \Omega^2_\rn (1+ x_\rp)^2
\mp 2 \Omega_\rn k_\parallel w_\parallel x_{\rp}  \right ]^{1/2}
\label{pair1_s}
\end{multline}
\begin{multline}
\sigma_{3,4} \approx \pm \Omega_\rn \left ( \frac{1}{x_\rp} - 1 \right ) - k_\parallel w_\parallel
\\
+ \frac{1}{x_\rp}
\left [ \Omega^2_\rn (1+ x_\rp)^2 \mp 2 \Omega_\rn k_\parallel w_\parallel x_{\rp}  \right ]^{1/2}
\label{pair2_s}
\end{multline}
From these results we see that the waves are unstable if
\be
| w_\parallel | > \frac{\Omega_\rn (1+ x_\rp)^2}{2 k_\parallel x_{\rm p}}
\label{crit2}
\ee
This criterion is clearly different from the weak drag result, (\ref{crit}).
Since $x_\rp$ is expected to be small, the critical relative flow needs to be larger
to trigger an instability in the strong drag case. However, the unstable modes no
longer depend on the mutual friction coefficient $\cal B$, as they did in the weak
drag case. Instead, the key coupling that facilitates the instability is
 due to ${\cal B}^\prime \approx 1$.

It is relevant to make an observation at this point. While the weak drag instability is well
known from studies of superfluid Helium, see \citet{glab,sidery07} for discussions, we believe
that our work provides the first suggestion that an analogous instability may operate in the strong drag problem.
Hence, the above results are exciting. Having said that, we can no longer
rely on the analogy with the Helium case. In the weak drag situation there is a natural
link between the instability and the transition to superfluid turbulence \citep{turbulent}.
In the strong drag case, it may seem natural to expect that the unstable modes
still lead to the formation of vortex tangles. However, given the stronger
coupling between the two fluids in the system the problem is different, and
will require further analysis in order to be understood.


\subsubsection{Adding shear viscosity}

The previous results were derived assuming vanishing shear viscosity, for the purpose of
understanding the properties of the various types of inertial waves.
However, viscosity could play an important role in the problem
due to the small scales involved. At the end of the day, if an instability is triggered in the inertial
waves shear viscosity should be the main counteracting agent and in order to arrive at
realistic estimates for the growth time of the instability we need to account for it.
Having said that, it is important to bear in mind  that the various inertial waves
will be affected  by viscosity in drastically different ways.
The key point is that only the proton fluid is directly coupled to the mechanism that
generates viscosity (recall that protons, electrons and the magnetic fluxtubes
are all assumed to be ``locked'' together) while any communication with the neutron component is
mediated by the mutual friction. This feature is clearly encoded in the hydrodynamical
equations (\ref{eulern})-(\ref{eulerp}).

The interplay between mutual friction and viscous coupling is exposed if we amend
the approximate mode solutions from the preceding Section with viscosity.
Considering first the short wavelength limit we now find
\begin{multline}
\sigma_{1,2} \approx \pm 2\Omega_\rn + (i{\cal B} \mp {\cal B}^\prime )
(2\Omega_\rn \mp k_\parallel w_\parallel ) \\
\pm
\frac{2i\Omega_\rn w_\parallel}{\nu_{\rm ee} k_\parallel x_\rp} \left[ {\cal B}^2 -({\cal B}^\prime)^2
\right]
\label{visc1}
\end{multline}
\begin{multline} 
\sigma_{3,4} \approx \pm (\Omega_\rn + \Omega_\rp^\parallel) + k_\parallel w_\parallel
\\
+\frac{2 \Omega_\rn}{x_\rp} (  i{\cal B}    
\mp {\cal B}^\prime )  + i \nu_{\rm ee} k_\parallel^2
\label{visc2}
\end{multline}
where in each case we have kept only the leading order viscous term\footnote{It is worth noting that
these results are obtained by holding $\nu_{\rm ee}$ fixed and then taking the large $k_\parallel$ limit. 
In effect, this means that one cannot use these expressions to deduce the $\nu_{\rm ee}\to 0$ behaviour.}. 
The key result is that
only the ``proton'' modes $\sigma_{3,4} $ are damped by viscosity in the usual way, that is,
via a  $i\nu_{\rm ee} k_\parallel^2$ term which dominates at small scales.
In contrast, the viscosity dependence of the ``neutron'' modes $\sigma_{1,2} $ is intimately linked to the
mutual friction coupling, leading to a perhaps surprising scaling with negative powers of $k_\parallel $.

As in the inviscid problem, the ${\cal R} \to 0 $ solutions simply follow from (\ref{visc1})
and (\ref{visc2}) after setting ${\cal B}= {\cal B}^\prime = 0 $. In this limit the $\sigma_{1,2} $
modes are essentially not affected by viscosity. This is not surprising since 
they describe oscillations in the neutron fluid only.

In the strong drag limit we find the following viscous modes
(assuming $\Omega^\parallel_\rp \approx \Omega_\rn $), which generalise (\ref{pair1_s}) and
(\ref{pair2_s}),
\begin{multline}
\sigma_{1,2} \approx \pm \Omega_\rn \left ( 1-\frac{1}{x_\rp} \right )
+ k_\parallel w_\parallel + \frac{i}{2} \nu_{\rm ee} k^2_\parallel \\
- \frac{1}{x_\rp} \Big[-\frac{1}{4}(\nu_{\rm ee} k^2_\parallel x_\rp )^2 \pm i x_\rp (x_\rp -1)
\Omega_\rn \nu_{\rm ee} k^2_\parallel \\ 
+  \Omega_\rn^2 (1+x_\rp)^2
\mp 2 \Omega_\rn k_\parallel w_\parallel x_\rp  \Big]^{1/2}
\label{visc3}
\end{multline}
\begin{multline}
 \sigma_{3,4} \approx \pm \Omega_\rn \left ( 1-\frac{1}{x_\rp} \right )
+ k_\parallel w_\parallel + \frac{i}{2} \nu_{\rm ee} k^2_\parallel \\
+ \frac{1}{x_\rp} \Big[-\frac{1}{4}(\nu_{\rm ee} k^2_\parallel x_\rp )^2 \pm i x_\rp (x_\rp -1)
\Omega_\rn \nu_{\rm ee} k^2_\parallel \\
+  \Omega_\rn^2 (1+x_\rp)^2
\mp 2 \Omega_\rn k_\parallel w_\parallel x_\rp  \Big]^{1/2}
\label{visc4}
\end{multline}
Not surprisingly these expressions admit an instability at leading order (top pair), but they also
demonstrate the contrasting viscosity dependence. To see this, consider the first pair of solutions
and assume that the viscosity is negligible. Then increase $\nu_{\rm ee} k_\parallel^2$ while keeping all other
parameters fixed. It is then easy to see that, once the leading viscous term under the square-root becomes dominant, there
will be a cancellation with the other $i\nu_{\rm ee} k_\parallel^2/2$ term. Thus, the leading viscous damping term does not
have the expected form, it is much weaker. 
This is in contrast to the result for the second pair of solutions. In that case the 
two viscosity terms combine to give the anticipated $i\nu_{\rm ee} k_\parallel^2$ damping term.
 
All these approximate viscous solutions convey the same physical information. The modes
that represent pure neutron inertial waves in the limit of vanishing mutual friction exhibit
an unconventional -- and much weaker -- dependence on shear viscosity. These are also the {\it only}
modes that can become unstable and lead to the formation of vortex tangles and turbulence.


\subsection{Growth timescales}
\label{sec:grow}

The numerical solution of the full system (\ref{equ1})-(\ref{equ4}) unveils the presence
of an instability for a large portion of the parameter space and, as we show below, the associated growth time
$\tau_{\rm grow}= 1/\mbox{Im}\ \sigma$ can be quite short. The numerical timescales can be well approximated by the
preceding analytic mode solutions (for different  wavelength ranges). Hence, we are able to provide simple analytic 
estimates for $\tau_{\rm grow}$. These estimates will be a useful tool in our  discussion 
of precessing neutron stars.

The first estimate follows from the short wavelength results,
(\ref{inv1}) or (\ref{visc1}),
\be
\tau_{\rm grow} \approx \frac{1}{2\Omega_\rn {\cal B}} \left
( \frac{\pi| w_\parallel|}{\Omega_\rn \lambda}  -1  \right )^{-1}
\label{grow}
\ee
Based on our previous discussion we expect this result to be  accurate also in the
weak drag limit, irrespective of wavelength (provided of course that $\lambda \ll R$).

A second estimate is relevant in the strong coupling regime. Since the relevant mode solutions are
sensitive to viscosity, we need to distinguish between the inviscid and the
viscous case. For the former we find from (\ref{pair1_s}),
\be
 \tau_{\rm grow} \approx \frac{x_\rp(1-x_\rp)}{\Omega_\rn}\,
\left  (\frac{4\pi |w_\parallel | x_\rp}{\Omega_\rn \lambda} -1 \right  )^{-1/2}
\label{grow_strong}
\ee 
In the viscous case it is not easy to extract a
similarly simple estimate. This is not surprising since the modes (\ref{visc3}) are more
complicated than the inviscid ones. Hence, we calculate
$1/\mbox{Im }\sigma $ numerically in this case.

\begin{figure}
\centerline{\includegraphics[height= 6.5cm,clip]{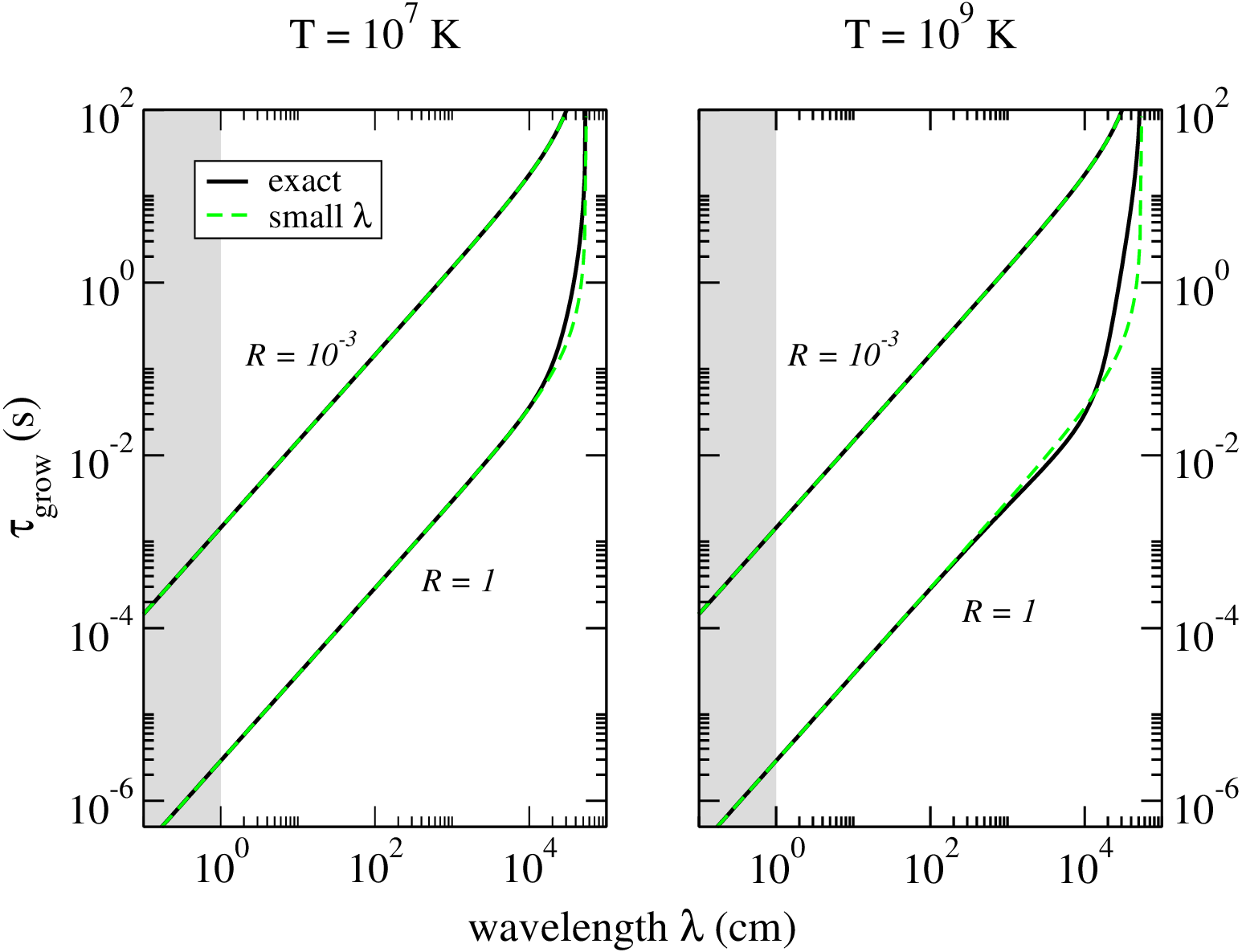}}
\caption{Numerical results (solid lines) for the instability growth time $\tau_{\rm
    grow} $ as a function of the wavelength $\lambda $, for two choices
  for the core temperature, $T = 10^7$~K and  $10^9~\mbox{K}$ (left and right panel, respectively). 
  The 
  analytic short wavelength approximation discussed in the main text is superimposed (dashed lines).
  The results represent weak ($\mathcal{R}=10^{-3}$) and intermediate ($\mathcal{R}=1$) drag. 
  The angle between $\vec{\Omega}_\rn $ and
  $\vec{\Omega}_\rp $ is taken to be  $1^\circ$ and the rotational period is $ P =
  1~\mbox{s} $. The  short-wavelength cut-off, which is  about 1~cm for this rotation rate, 
is indicated by a grey area.} 
\label{fig1}
\end{figure}

\begin{figure}
\centerline{\includegraphics[height= 6.5cm,clip]{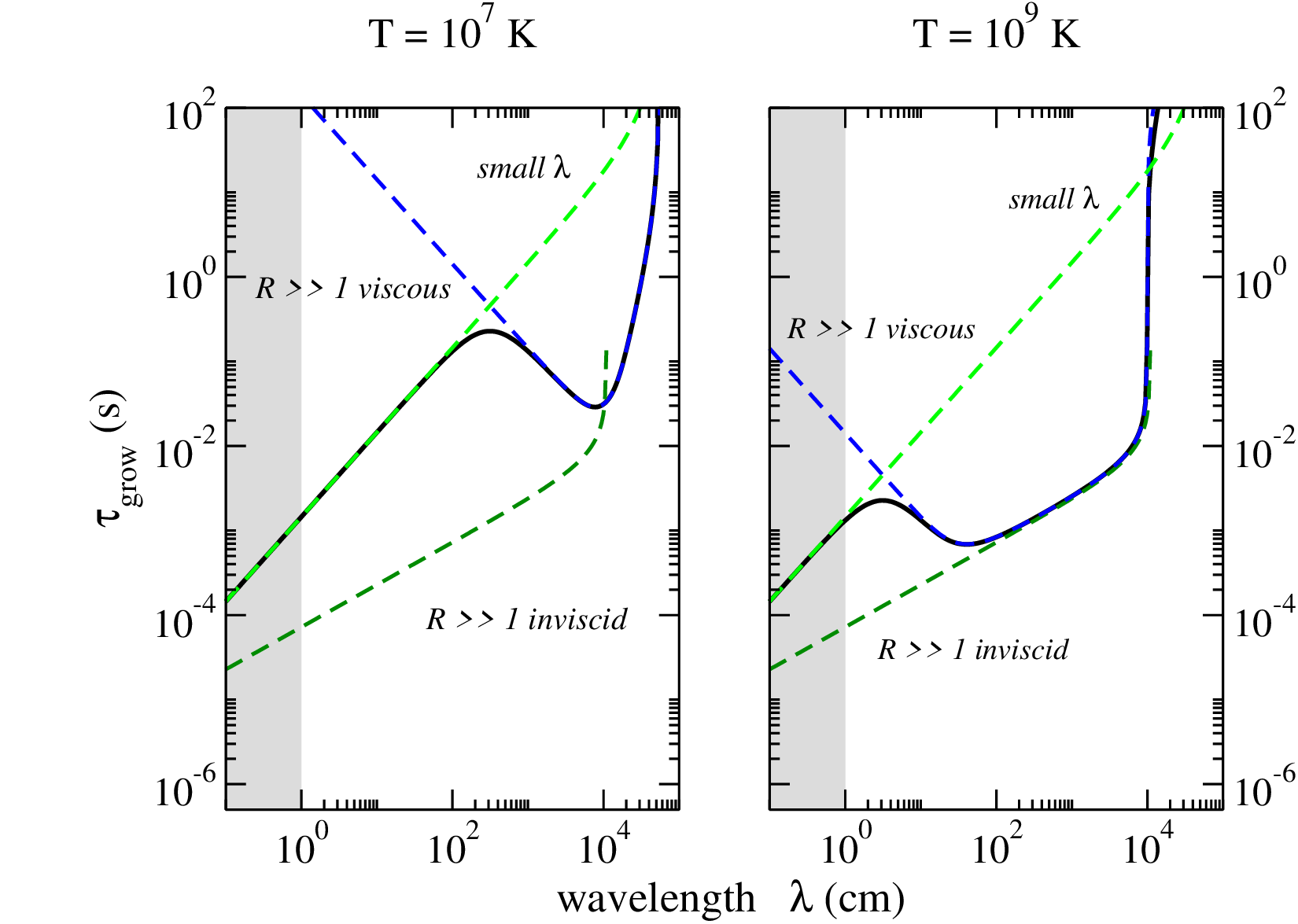}}
\caption{Same as Figure~\ref{fig1}, but for strong drag ${\cal R}=10^3 $. In this regime, shear viscosity affects the results 
for intermediate wavelengths (around $\lambda = 1000$~cm and 10~cm in the 
left and right panels, respectively.)  The 
  various approximations discussed in the main text are superimposed (dashed lines).
It is important to note that the unstable modes are not 
strongly affected by shear viscosity in the  
short wavelength limit. }
\label{fig1b}
\end{figure}

\begin{figure}
\centerline{\includegraphics[height= 6.5cm,clip]{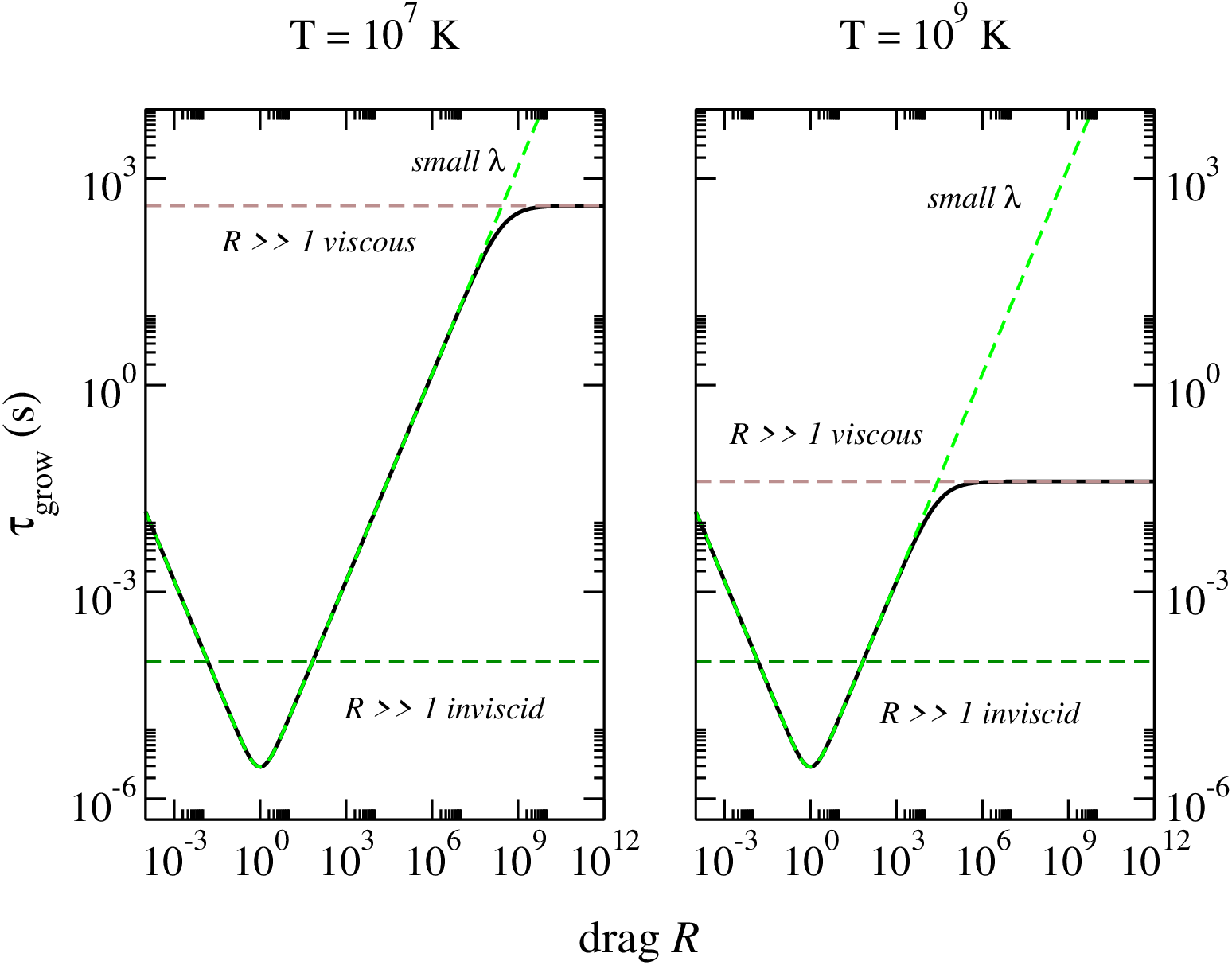}}
\caption{Numerical results (solid lines) for the instability growth time $\tau_{\rm
    grow} $ as a function of mutual friction drag ${\cal R}$ for two
  choices of core temperature, $T = 10^7$~K and  $10^9~\mbox{K}$ (left and right panel, respectively).  The
  analytic approximations discussed in the main text are superimposed (dashed lines).
  Here we have taken the wavelength to be that of the short wavelength cut-off, i.e.
 for $ P = 1~\mbox{s} $ we have  $\lambda = 1~\mbox{cm} $.  The angle between
  $\vec{\Omega}_\rn $ and $\vec{\Omega}_\rp $ is assumed to be $1^\circ$}
\label{fig2}
\end{figure}

\begin{figure}
\centerline{\includegraphics[height= 6.5cm,clip]{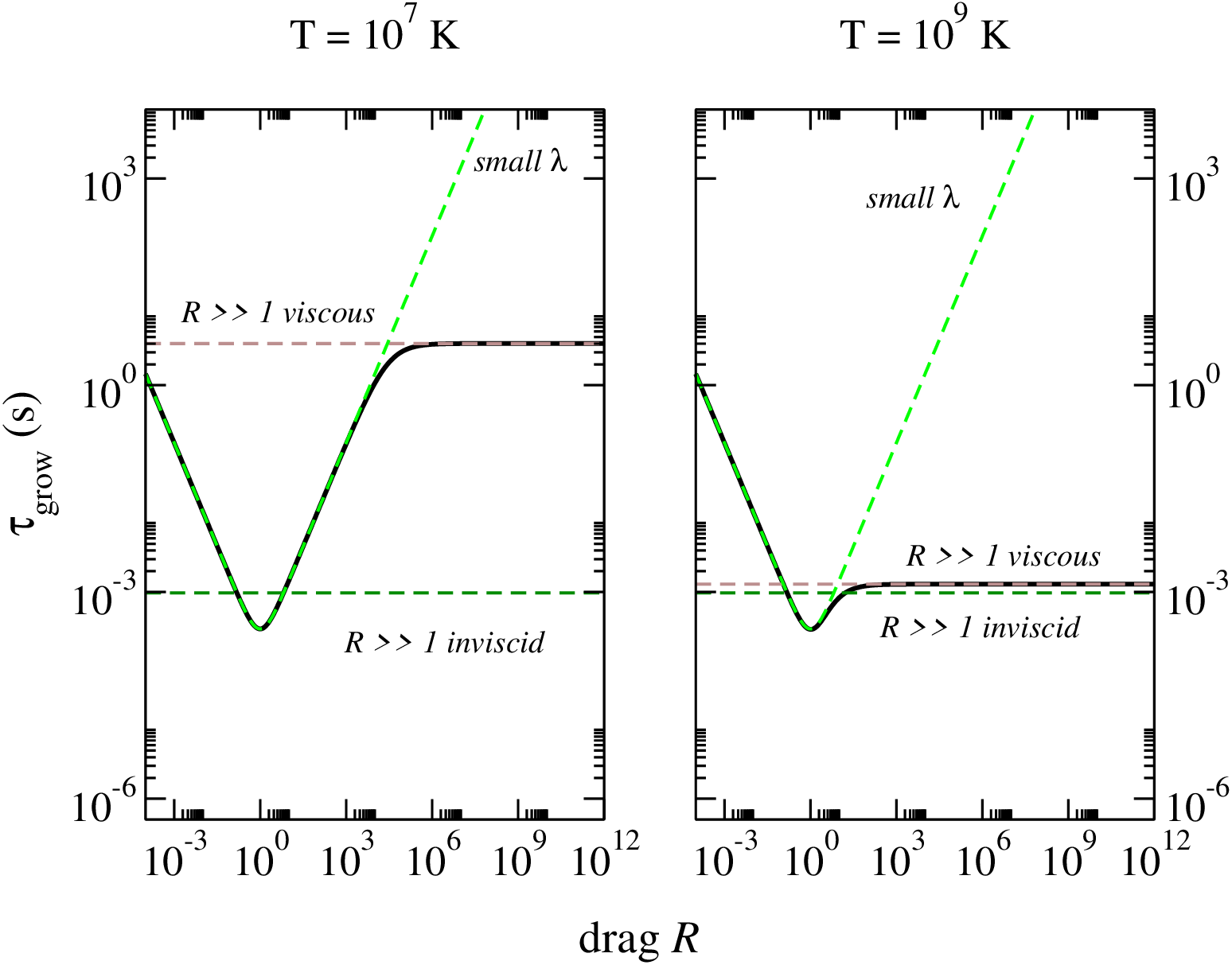}}
\caption{Same as Figure~\ref{fig2}, but for  $\lambda = 100~\mbox{cm} $.}
\label{fig2b}
\end{figure}

Our results for the instability growth times are illustrated in Figures~\ref{fig1}-\ref{fig3}. 
In the figures we compare the different approximations for $\tau_{\rm grow} $ 
to the full numerical results. The data in the figures are obtained using a $w_\parallel$ 
corresponding to a mismatch between $\Omega_\rp^i$ and $\Omega_\rn^i$ of $1^\circ$
and a rotation period of 1~s. Later we will relate these results to actual 
precession solutions.

Let us first consider Figure~\ref{fig1}, where we show the growth time as a function of the wavelength
for fixed temperature and drag coefficient. The data concerns the intermediate and weak drag regimes, 
explicitly  ${\cal R} = 1 $ and  ${\cal R} = 10^{-3} $. As we discussed earlier, there will be 
a short wavelength cut-off. A more detailed analysis would link this to the
validity of the averaging that led to the macroscopic hydrodynamics equations
(\ref{eulern}) and (\ref{eulerp}). This is a very difficult problem.
To make immediate progress, we will simply assume that the instability analysis
becomes invalid once the wavelength is so short that the fluid description is no longer valid.
Then it seems reasonable to use something like
\be
\lambda_\mathrm{min} \approx 100 d_\rn \approx \left( {P \over 1 \mathrm{s} } \right)^{1/2} \ \mathrm{cm}
\ee
The expected short-wavelength cut-off
for a rotation period $P=1$~s is then about 1~cm, as indicated in the figure. From these results it
is clear that the short wavelength approximation (\ref{grow}) is
very good for (essentially) the entire
$\lambda$ range and all relevant temperatures.

In the strong drag regime, for ${\cal R} \gg 1 $,   the
variation of growth time with wavelength is more complicated, cf. Figure~\ref{fig1b} which 
shows data for $\mathcal{R}=10^{3}$. 
The small $\lambda $ approximation again matches the exact result well in the region near
the short wavelength cut-off, but it is not accurate for all wavelengths. 
For intermediate 
wavelengths the strong drag viscous solution (\ref{visc3}) is more
suitable. By combining the two approximations we arrive at an almost perfect agreement with the
numerical results, including the transition region where shear viscosity is relevant.  
It is also apparent from the figure that the inviscid estimate
(\ref{grow_strong}) is accurate only for relatively hot
stars, for which shear viscosity is weak. This is as expected.

We get a different view of the results if we fix the wavelength and 
vary the drag coefficient $\mathcal{R}$. Figure~\ref{fig2} shows such results for 
a rotation period of $P=1$~s at the corresponding short wavelength cut-off, i.e. 
$\lambda=1$~cm. From the left panel we see that the small $\lambda $ approximation 
(\ref{grow}) is accurate for ${\cal R} < 10^7 $ or so at $T=10^7$~K. The remaining parameter
space is well approximated by the viscous ${\cal R} \to \infty $ solution
(\ref{visc1}). The inviscid strong-drag approximation (\ref{grow_strong}) is never really 
adequate for this very short wavelength. 
Similar results for $\lambda=100$~cm are shown in Figure~\ref{fig2b}. Note that in 
both figures the growth time is shortest for $\cal R$ of order unity, and the 
general behaviour follows immediately from $\tau_\mathrm{grow}\sim 1/\mathcal{B} \sim \mathcal{R} + 1/\mathcal{R}$.
This is analogous to the fact that the glitch relaxation timescale is expected to be long in both 
the weak and the strong drag limit \citep{alpar}.

One can show that   
for very strong coupling, ${\cal R} > 10^5 $ or so, the fastest growing waves
are those with moderately short wavelengths, $\lambda > 100~\mbox{cm}
$.  For such large values of ${\cal R} $ the mode solution
(\ref{visc3}) is accurate even for very short wavelengths. Whether this very strong drag regime is 
physically relevant is not clear. 

From these comparisons with the numerical results we arrive at
a practical guide to the validity of the analytic timescale
estimates. For a wide range of $\mathcal{R}$ the fastest growing modes are  
in the short wavelength regime near the cut-off, i.e. $\lambda \sim 1~\mbox{cm} $ for $P=1$~s, where the
estimate (\ref{grow}) applies.  For very strong coupling (depending on the temperature)
the instability grows faster in modes with moderately short
wavelengths (above $\lambda \sim 100~\mbox{cm}$) in which case
(\ref{visc3}) applies. In fact, this approximation tends to be good for 
intermediate wavelengths at all temperatures for strong drag.


\section{Are precessing neutron stars stable?}

In order to address this key question we first need to discuss the ``standard'' two-fluid
model for neutron star precession 
\citep{swc} in some detail.  A self-contained discussion of the relevant precession
modes can be found in Appendix~\ref{app:precess}. We want to
combine these precession solutions with the preceding plane-wave analysis to 
establish whether the superfluid instability is present or not. In
principle, this analysis may provide us with a ``no-go theorem'' for
neutron star precession. After all, the precession modes are obtained
by assuming global solid body rotation. If the corresponding solution
exhibits a rapidly growing short wavelength instability, then the
assumption of solid body rotation does not hold.  The precession
solution would simply be inconsistent. The upshot of this would be
that one cannot neglect the hydrodynamical degrees of freedom of the
precession problem. Any analysis 
of the precession problem would have to
account for the hydrodynamics of the neutron star core, which would be a
serious computational challenge.  

In our analysis of the precession modes in Appendix~\ref{app:precess}
we go beyond the original work of \citet{swc} by providing
general expressions for the precession period $P_{\rm pr} $, the damping
time $\tau_d $ and the angular amplitude $\theta_w $ (the so-called
``wobble'' angle) for arbitrary mutual friction coupling.  These are
the key parameters\footnote{Note that $P_{\rm pr} $ and $\theta_w $
  are in fact observable quantities, extracted from the timing data of
  the candidate precessors \citep{dij,link01}.} which enable us to make contact with the
two-fluid instability analysis.  However, it turns out that the range
of ${\cal R} $ for which precession is not overdamped by mutual
friction (in the sense that $\tau_d > P_{\rm pr} $ ) is entirely
covered by the approximate (weak and strong drag) results of
\citet{swc}. This is clearly illustrated in Figure~\ref{overdamp}.
Therefore we only discuss results for the
stability of precession in the limits of weak and strong drag.

From the practical point of view, the
perturbative calculation in Section~\ref{sec:local} assumes a static
background. Therefore a given precession mode can only serve as 
background provided that the unstable waves grow fast enough. To ensure that this is the case, we 
we will require $\tau_{\rm grow} \ll \mathrm{min}(P_{\rm pr},\tau_d)$ for the
instability to be astrophysically relevant in a precessing star.


\subsection{Weak drag precession}
\label{sec:weak}

As described in Appendix~\ref{app:precess}, there are two precession modes for any given value of the
mutual friction parameter $\mathcal{R}$.
Let us first consider the case of weak drag precession ($\mathcal{R} \ll 1$) and
the mode that, in this weak coupling limit, reduces to the
standard classical free precession of a biaxial body (i.e. to the sort
of precession commonly considered in the literature, see e.g.
\cite{dij,link01}).  This mode represents weakly
damped, long-period precession with 
\be 
P_\mathrm{pr} = \frac{P}{\epsilon} 
\ee 
where $\epsilon $ is the (dimensionless) deformation associated with
the biaxiality in the crust's moment of inertia (see
Appendix~\ref{app:precess}). For PSR B1828-11 this deformation is
estimated to be $\epsilon \approx 10^{-8}$.
Moreover, assuming that PSR B1828-11 is described by the
weak drag precession solution, one should be able to constrain ${\cal B}$ using the fact
that since its discovery \citep{1828} several undamped precession oscillations
have been monitored. If $N$ is the number of oscillations then we should have $\tau_d > N P_{\rm pr} $ 
from which we find ${\cal R} < (0.16/N) (I_1/I_s) $. Using $N\approx 10 $ and $I_1/I_s \approx 0.1 $
we find ${\cal R} < 2 \times 10^{-3}$. Combining this upper limit with the lower limit
placed by Vela's glitch relaxation timescale we find that only a relatively narrow interval
of weak drag mutual friction is allowed, $ 2\times 10^{-5}  < {\cal R} <  2 \times 10^{-3}$.
Interestingly, the  value suggested by \citet{als,sidery06}, ${\cal R} \sim 10^{-4} $, falls within this range.

The weak drag instability criterion (\ref{crit}) implies that waves
with wavelength shorter than 
\be \lambda_\mathrm{max} = { \pi
  |w_\parallel | \over \Omega_\rn} \ee 
are unstable. For the given
weak drag precession solution we have (Appendix~\ref{app:precess})
\be w_\parallel \leq \epsilon
\theta_w \Omega R
\label{w_weak}
\ee
which means that 
\begin{multline}
\lambda_\mathrm{max}  \leq    \pi \epsilon \theta_w R \\
\approx
5.4\times 10^{-4} \left (\frac{\theta_w}{1^\circ} \right ) \left( {\epsilon \over 10^{-8} } \right) 
\left( { R \over 10^6 \mathrm{cm} } \right) \ \mathrm{cm}
\end{multline}
Since we need to have $\lambda_\mathrm{max}>\lambda_\mathrm{min}$ in order to argue that the instability is
relevant we see that we must have
\be
\left(\frac{\theta_w}{1^\circ} \right ) > 1900 \left( {P \over 1 \mathrm{s} } \right)^{1/2}
\left( {\epsilon \over 10^{-8} } \right)^{-1} \left( { R \over 10^6 \mathrm{cm} } \right)^{-1}
\label{weak_wobble}
\ee 
In words, this inequality gives the wobble angle required for the
fluid in at least some portion of the star to be unstable for 
 inertial waves with $\lambda = \lambda_{\rm min}$. 

To demonstrate that the instability analysis is consistent we need to
show that the growth time for the unstable modes is short compared to
the precession period.  Using (\ref{grow}), which according to Figure~\ref{fig1} matches the
exact weak drag numerical results, we find
\be
\tau_\mathrm{grow} \approx  { \lambda \over 2 \pi {\cal B} |w_\parallel|}
\label{karlaxel}
\ee
This should be a good approximation for the shortest wavelength modes in the problem. Using the relevant 
result (\ref{w_weak}) for $w_\parallel$ we find that
\begin{multline}
\tau_\mathrm{grow} > \\
4 \times 10^{11} \left( { {\cal B} \over 4 \times 10^{-4} } \right)^{-1}
\left( {\epsilon \over 10^{-8} } \right)^{-1}
\left( {\theta_w \over 1^\circ} \right)^{-1} \left( { P \over 1\ \mathrm{s} } \right)
\left( { \lambda \over R } \right)\ \mathrm{s}
\end{multline}
On the other hand, the precession period is
\be
P_{\rm pr} \approx 10^8  \left( {\epsilon \over 10^{-8} } \right)^{-1}
\left( { P \over 1\ \mathrm{s} } \right) \ \mathrm{s}
\ee
Requiring that $\tau_\mathrm{grow} < 0.1 P_\mathrm{pr}$ we see that we must have
\be
\lambda < 25  \left( { {\cal B} \over 4 \times 10^{-4} } \right)\left( {\theta_w \over 1^\circ} \right)
\left( { R \over 10^6 \ \mathrm{cm}} \right) \ \mathrm{cm}
\ee
We need to ask whether there is room for  unstable waves
with wavelengths longer than $\lambda_\mathrm{min}$. This would be the case if 
\be
\left(\frac{\theta_w}{1^\circ}\right) > 4 \times 10^{-2}  \left( { {\cal B} \over 4 \times 10^{-4} } \right)^{-1}
\left( { R \over 10^6 \ \mathrm{cm}} \right)^{-1} \left (\frac{P}{1~\mbox{s}} \right )^{1/2}
\ee
In words, this inequality gives the wobble angle required for the
unstable $\lambda = \lambda_{\rm min}$ inertial waves in some portion
of the star to have a growth time of less than a tenth of the free precession
period.  Compared to (\ref{weak_wobble}) this
restriction on the wobble angle is clearly much less stringent. Hence, we learn that
whenever the instability is active the growth time
$\tau_{\rm grow} $ is much shorter than the precession period.

If we associate the instability with the onset of superfluid
turbulence (see \citet{turbulent}) then only stars that satisfy the
criterion (\ref{weak_wobble}) are likely to be turbulent.  Note that
this hardly constrains the realistic parameter space since according
to the data all candidate precessors have $\epsilon \sim
10^{-7}-10^{-8} $. Hence it would seem safe to conclude that this
instability does not play a role in systems where the drag is weak,
unless they are significantly deformed.  However this does not mean
that the instability criterion (\ref{weak_wobble}) is not
astrophysically relevant.  For example, a millisecond pulsar with
(say) $P=1.5~\mbox{ms}$ and $\epsilon = 5 \times10^{-7}$ (which is a
realistic value) would undergo an instability for a wobble angle
$\theta_w > 1.5^\circ $. A more extreme example would be 
the suggested sighting of free precession of the SN1987A remnant \citep{mid1,mid2}.
Taking a spin period of 2.14~ms and a free precession period of $\sim 10^3$~s, 
we would have $\epsilon \approx 2\times 10^{-6}$.  For these values, 
the instability criterion  (\ref{weak_wobble}) would be satifisfied for
$\theta_w > 0.4^\circ$. As discussed by \citet{dij}, the wobble angle would have to be much larger than 
$1^\circ$ in order to explain the observations. In other words, this system would 
suffer the superfluid instability.    

Now consider the second mode in the weak coupling limit.  As discussed
in Appendix~\ref{app:precess}, in the case of zero coupling this mode
simply corresponds to a misalignment between the rotation axes of the
neutron and proton fluids, with both angular velocity vectors fixed in
the inertial frame, i.e. no precession.  Weak coupling then serves to
induce a slow precession, with the star rotating about the 3-axis of
the biaxial crust, and with this axis itself tracing out a cone at the
longer free precession period.  This is the opposite of the standard
free precession motion of the first mode, where the 3-axis traces out
a cone in space at the ``spin'' frequency, with a slow additional
precession rotation superimposed.  

However, from the observational point of view, both motions would
produce modulations in the radio data and so this second mode cannot
be discounted.  A crucial difference is that for this second mode the
ratio $P_{\rm pr}/P$ is determined not by the biaxiality of the crust
$\epsilon$, but instead by the strength of the coupling. We have
\be 
\frac{P}{P_{\rm pr}} \approx \mathcal{R} \frac{I_1}{I_s + I_1} 
\ee
The observed $P/P_{\rm pr} \sim 10^{-8}$ would then give an estimate
of ${\cal R} \sim 10^{-7}$ or less.  This is smaller than the typical
estimates based on microphysical considerations (see for example \citet{sidery06}) and
is also smaller than the lower bound placed on $\cal R$ from glitch
observations (see discussion at the end of
Section~\ref{sec:form}), arguing against this mode as an explanation of
the PSR B1828-11 observations.

A further argument against this mode being  relevant comes
from our stability analysis.  For this mode we find that
\be
w_\parallel \le \tilde{I} \theta_w \Omega R
\ee
and
the instability is
present for wavelengths shorter than 
\be\lambda_\mathrm{max} = \pi \tilde{I} \theta_w R
\ee
where 
\be
\tilde{I} = 1 + { I_1 \over I_s} 
\ee
This means that, if we require $ \lambda_\mathrm{max}>\lambda_\mathrm{min}$ then an instability is present if
\be
 \left( {\theta_w \over 1^\circ} \right)  > 
2\times 10^{-5}  { 1 \over \tilde{I}} \left( {P \over 1 \mathrm{s}} \right)^{1/2} \left( { R \over 10^6 \ \mathrm{cm}} \right)^{-1}
\ee
Alternatively, this can be expressed as a constraint on the rotation period. To avoid the instability we require
\be
P > 3\times 10^9  \tilde{I}^2  \left( {\theta_w \over 1^\circ} \right)^2 \left( { R \over 10^6 \ \mathrm{cm}} \right)^2 \ \mathrm{s}
\ee
As one would have expected given the large relative flow associated with it,  
this mode is seriously affected by the superfluid instability. The growth time is 
\be
\tau_\mathrm{grow} > 3.6\times10^3 { 1 \over \tilde{I}} \left( {P \over 1 \mathrm{s}} \right)
 \left( { {\cal B} \over 4 \times 10^{-4} } \right)^{-1}  \left( {\theta_w \over 1^\circ} \right)^{-1} 
\left( { \lambda \over R } \right) \ \mathrm{s}
\ee
If we require that this is smaller than a tenth of the precession period, then we must have 
\be
\lambda < 3\times10^2 \tilde{I}  \left( { {\cal B} \over 4 \times 10^{-4} } \right) \left( {\theta_w \over 1^\circ} \right)\ \mathrm{ cm} 
\ee
Finally, requiring this to be larger than the cut-off $\lambda_\mathrm{min}$ we see that the instability is 
present for
\be
\left( { \theta_w \over 1^\circ} \right) > 3 \times 10^{-3} { 1 \over \tilde{I} }\left( { {\cal B} \over 4 \times 10^{-4} } \right)^{-1}
\left( {P \over 1 \mathrm{s}} \right)^{1/2}
\ee 
The analysis 
suggests that this mode would be unstable for any significant misalignment. 
It is therefore unlikely to be present in real stars.


\subsection{Strong drag precession}
\label{sec:strong}

For strong  mutual friction the nature of the precessional motion  
changes significantly. To some extent, this was first discussed by \citet{shaham77}, see Appendix~\ref{app:precess}
for details. Precession is fast, with period   
\be
P_\mathrm{pr} \approx P \left ( \epsilon + \frac{1}{x_\rp} \right )^{-1} \approx x_\rp P
\label{Ppr_strong}
\ee

As we discussed in Section~\ref{sec:grow}, in the case of strong drag the instability is sensitive to
the action of viscosity and as a consequence the ``window'' of the fastest growing wavelengths shifts 
with varying ${\cal R} $. The most natural choice is to assume that $1 \ll {\cal R} < 10^5 $ (see discussion
at the end of Section~\ref{sec:form}) which 
means that the small $\lambda $ approximation (\ref{grow}) for $\tau_{\rm grow} $ describes  
the fastest growing waves accurately. We again have (\ref{karlaxel}),
which can be combined with the obtained approximation for $w_\parallel$ (Appendix~\ref{app:precess})
\be
w_\parallel \leq \frac{\theta_w \Omega R}{x_\rp}
\label{w_strong}
\ee
For typical parameters we find 
\be
\tau_\mathrm{grow} > 140  \left( { x_\rp \over 0.1} \right)
\left( {\theta_w \over 1^\circ } \right)^{-1} \left( { P \over 1\ \mathrm{s} } \right) 
\left( { \mathcal{R} \over 10^3 } \right) \left( { \lambda \over R } \right) \ \mathrm{s}
\ee
For consistency $\tau_\mathrm{grow} $  needs to be shorter than the precession period (\ref{Ppr_strong}).
Requiring $\tau_\mathrm{grow} < 0.1P_{\rm pr}$ as before, this restricts the wavelength to
\be
\lambda < 70 \left( {\mathcal{R} \over 10^3} \right)^{-1} \left( {\theta_w \over 1^\circ } \right) \ \mathrm{cm}
\ee
The short lengthscale cut-off obviously remains as in the weak drag case.
Thus we see that we need
\be
\left (\frac{\theta_w}{1^\circ} \right ) > 1.4 \times 10^{-2}  \left( {\mathcal{R} \over 10^3} \right)
\left( { P \over 1\ \mathrm{s} } \right)^{1/2}  
\ee 
in order for the instability to be relevant. This clearly sets a very severe constraint on 
the fast precession solution.

\begin{figure}
\centerline{\includegraphics[height= 6.5cm,clip]{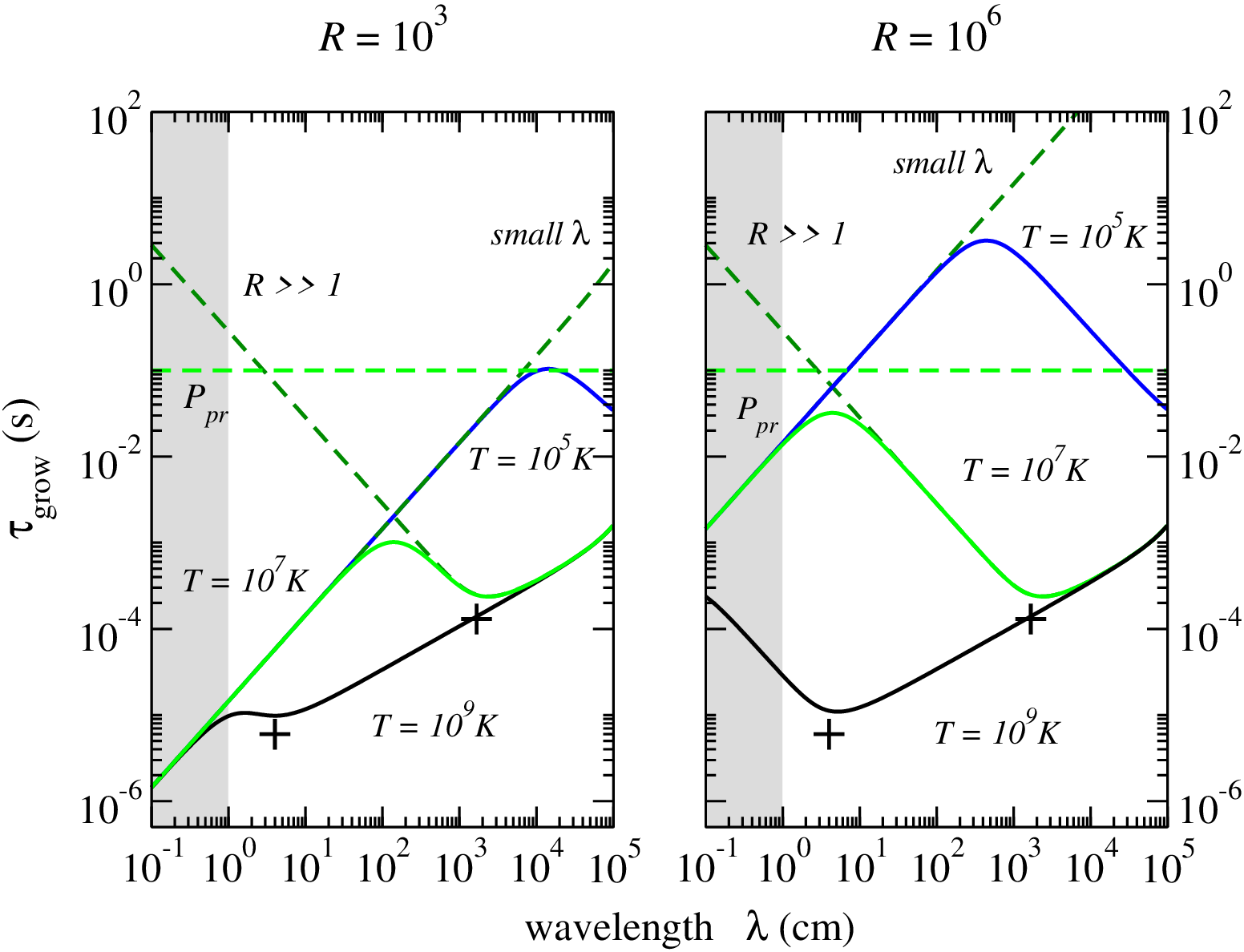}}
\caption{Instability growth times $\tau_{\rm grow}(\lambda)$ for strong drag fast precession with wobble angle
$\theta_w = 1^\circ$. Two choices for
the drag coefficient are shown, ${\cal R} = 10^3$ and $10^6 $ (left and right panel, respectively). 
The instability is present for all wavelengths down to $\lambda_{\rm min} $ (shaded area), despite the action of
viscosity. For a wide range of core temperatures $\tau_{\rm grow}$ is well below the 
precession period $P_{\rm pr} \approx x_\rp P $ (we have used $P=1~\mbox{s}$  and  $x_\rp=0.1$) which is 
shown as a horizontal dashed line. The analytic approximations discussed in the main text are 
superimposed as dashed curves for the $T=10^7~\mbox{K}$ case. Estimates for the minima in the $T=10^7$~K and 
$10^9$~K curves, obtained from (\ref{lamex}) and (\ref{tauex}) are indicated by $+$ symbols.}
\label{fig3}
\end{figure}

The above results are, however, only relevant for short wavelengths. 
It is, of course, straightforward to find the general solution numerically. 
This provides the exact $\tau_{\rm grow}$ as shown in 
Figure~\ref{fig3}. The numerical timescales  corroborate the  analytic estimates.
For a large part of the parameter space the instability grows at sub-millisecond timescales, much faster
than the precession period. This would be the case, for example, for PSR B1828-11 ($P \approx 0.4~\mbox{s},~
T \sim 10^7~\mbox{K}$) if it were to precess under strong mutual friction. 

As we have already discussed, the strong-drag results are temperature dependent. 
This is clear from the results shown in Figure~\ref{fig3}.
For very large ${\cal R}$, the fastest growing instability is no longer associated with the short-wavelength cut-off region. 
Instead, the fastest $\tau_{\rm grow}$ is  temperature dependent. This longer wavelength part of the instability window 
can be understood from (\ref{visc3}) and (\ref{visc4}). For longer wavelengths (small $k_\parallel$) one would expect 
shear viscosity to be unimportant. Thus the modes of the system should be well approximated by the inviscid result. 
In the inviscid case (\ref{visc3}) implies that waves with wavelength shorter than
\be
\lambda_\mathrm{max} \approx { 4 \pi |w_\parallel| \over \Omega_\rn}
{ x_\rp \over (1+x_\rp)^2} \approx  { 4 \pi x_\rp |w_\parallel| \over \Omega_\rn}
\ee
will be unstable.
Working out the associated growth time we find 
\be
\tau_\mathrm{grow} > 3.4 \times 10^{-2} \left( { x_\rp \over 0.1} \right)
\left( {\theta_w \over 1^\circ} \right)^{-1/2} \left( { P \over 1\ \mathrm{s} } \right)
\left( { \lambda \over R } \right)^{1/2} \ \mathrm{cm}
\label{tginv}\ee
This estimate should be valid for long wavelengths  ($<\lambda_{\rm max}$) and/or high temperatures. As we move towards shorter wavelengths
at fixed $T$, the importance of shear viscosity will increase, cf. Figure~\ref{fig3}. To estimate the point at which the viscous
contribution becomes important we can balance the first and last terms under the square-root in (\ref{visc3}). 
That is, we consider
\be
{ 1 \over 4} \nu_{\re\re}^2 k_\parallel^4 x_\rp^2 = 2 \Omega_\rn k_\parallel w_\parallel x_\rp
\ee
This suggests that the viscosity starts to play a role when 
\be
\lambda \approx 360 \left({P\over 1 \mathrm{s}} \right)^{2/3}  \left({R\over 10\ \mathrm{km}} \right)^{-1/3}
\left({\theta_w\over 1^\circ} \right)^{-1/3} \left({x_\rp\over 0.1} \right)^{2/3}
\left({ T \over 10^8 \mathrm{K}} \right)^{-4/3}\ \mathrm{s}
\label{lamex}\ee
Combining this with the inviscid growth timescale (\ref{tginv}) we find
\be
\tau_\mathrm{grow} \approx 6 \times 10^{-4} \left({P\over 1 \mathrm{s}} \right)^{4/3} 
\left({\theta_w\over 1^\circ} \right)^{-2/3} \left({x_\rp\over 0.1} \right)^{4/3}
\left({ T \over 10^8 \mathrm{K}} \right)^{-2/3}\ \mathrm{s}
\label{tauex}\ee 
Adding the estimated ($\lambda, \tau_\mathrm{grow}$) points to the data in Figure~\ref{fig3} we see that we
have arrived at a reasonable approximation for the fastest growing mode in the intermediate wavelength 
regime. Finally, we can use this estimate to check when one would expect there to be such an instability window.
If we require  $\tau_\mathrm{grow} < 0.1 P_\mathrm{pr}$ as usual, then we find that the 
longer wavelength instability is present for
\be
T > 1.5 \times10^6 \left({P\over 1 \mathrm{s}} \right)^{1/2} 
\left({\theta_w\over 1^\circ} \right)^{-1} \left({x_\rp\over 0.1} \right)^{1/2}\ \mathrm{K}
\ee

To conclude, our results provide a strong indication that the fast
precession solution may not be relevant for  
realistic systems. Since it is generically unstable to
small scale inertial waves the solid-body rotation assumption does not
hold and the solution is inconsistent.  


\section{Summary and outlook}

Within the standard two-fluid model for superfluid neutron stars we have discussed 
a new class of instabilities
that set in through short wavelength inertial waves, provided there is sufficient 
relative flow along the neutron vortices. As a demonstration of the astrophysical
relevance of these instabilities we have considered freely precessing neutron stars, 
 in which the two angular velocity vectors are naturally misaligned. Our analysis is 
based on the usual form for the superfluid mutual friction, which depends on the strength 
of the drag that determines the friction between neutron vortices and the proton/electron fluid.
Our results show 
that precession may trigger instabilities both in the weak- and strong drag regimes. 
Based on the analogy with the so-called Donnelly-Glaberson instability in superfluid 
Helium \citep{glab,turbook}, we
argue that this instability is likely to lead to the formation of vortex tangles and generate 
turbulence in superfluid neutron star interiors.

The standard model \citep{swc} predicts two possible 
 precession regimes. Weak drag leads to slowly damped long period precession, while strong drag 
implies that precession should be fast. 
Assessing the stability of each precession mode we conclude that slow precession 
is generally stable. In particular, this should be true for the few known candidate precessors, 
like PSR B1828-11. An interesting exception could be slow precession of millisecond-period 
neutron stars. These may become unstable above a wobble angle $\theta_w \sim 1.5^\circ $ assuming a 
crust deformation $\epsilon \sim 5 \times 10^{-7} $, well within the theoretically allowed range \citep{cutler,haskell}. 

In contrast, we find the fast precession solution to be unstable    
under  generic conditions. This suggests that the solid-body rotation assumption that 
leads to the precession mode \citep{swc} is inconsistent. The upshot of this could be that fast precession is not 
realised in an astrophysical system. It should certainly be the case that one cannot ignore the 
small scale hydrodynamical degrees of freedom in the neutron star core in an analysis of the
precession problem. This is an interesting conclusion since it adds significant complexity to the modelling. 
It also calls into question some well established ideas. 

Since the seminal discussion of \citet{shaham77}, it has been 
known that precession is sensitive to the neutron vortex dynamics. 
This is natural since the vortices provide the bulk of the star's rotation. 
If the vortices interact strongly with the charged component (e.g. the crust) then the
gyroscopic nature of the vortex array will inevitably lead to fast precession. 
This corresponding motion can either be described in terms of a ``pinned'' component
affecting the usual Eulerian precession \citep{dij} or as a strong drag fast precession solution in the 
framework of \citet{swc}. The observed long period precession of PSR B1828-11 is 
thus consistent only with weak mutual friction. 

However, one can argue that the interaction 
between neutron vortices and the (much more numerous) magnetic fluxtubes in a type~II proton 
superconductor ought to lead to strong drag \citep{sauls,ruderman}. If this leads to the core vortices being effectively pinned
to the charged component, then fast precession should result (as in the Shaham model). If on the other hand, 
the precession motion is able to force the vortices through the fluxtubes then the motion 
should be highly dissipative and there would be no lasting precession. This led \citet{link03} to the suggestion that 
the apparent conflict with the observations can be avoided if the protons instead form a type~I 
superconductor. There would then be no fluxtubes and slow precession may be possible.
However, this would not accord with the generally held view that the protons in the outer neutron star core will 
condense into a type~II superconductor.  

Our results add yet another twist to this story. We have demonstrated that the fast precession solution,  
that would result if the vortices and the type~II fluxtubes interact strongly, is generically unstable. 
Although we cannot claim to understand the dynamical repercussions of this, it seems reasonable to assume that the 
fast solid-body precession motion of the \citet{swc} model will be significantly affected. If the instability 
leads to a state of superfluid turbulence, as the analogy with the Helium problem suggests, then 
it could well be that the precession does not survive. To explore this, a meaningful analysis must 
venture beyond the rigid-body precession model and account for detailed
superfluid hydrodynamics. This may be prohibitively difficult. Anyway, it is clear that we cannot yet use 
the precession observations to draw any 
reliable conclusions concerning the state of the proton superconductor. 

The present work can (and needs to) be extended in a number of ways. Our uniform density neutron star model
is too simplistic and one would certainly want to account for interior stratification etcetera.
If one wants to consider superfluid-superconducting mixtures seriously then 
the two-fluid model probably needs to be upgraded into a more realistic setup with
 three fluids (after relaxing the assumption of comoving charged particles), plus the
magnetic field. The dynamics of such a model for neutron star cores 
is largely unexplored \citep{mendell98}, yet the additional degrees of freedom could
lead to new features.   
Our calculation was perturbative. Hence, little can be said about the non-linear
behaviour of the unstable waves. Most likely this will be complicated, due to 
the small scales involved. Numerical simulations of this problem may be challenging, but 
should be encouraged. In fact they may be required if we are to understand the fate of a 
precessing neutron star in the strong-drag regime. Some of the necessary tools 
have already been developed by \citet{peralta05,peralta06}, and there is analogous work 
for superfluid Helium \citep{turbook}.
Finally, at the level of linearised theory the present two-fluid analysis could be modified 
to consider the inner layers of the crust, where superfluid neutrons 
coexist with an nuclear lattice. One may expect the nature of any superfluid instabilities in that
region to be different. 

These are all interesting problems that we expect to return to in 
the future.


\section*{Acknowledgements}

This work was supported by 
PPARC/STFC via grant number PP/E001025/1. NA also acknowledges support from PPARC via Senior Research Fellowship no
PP/C505791/1.


\appendix

\section{Explicit precession solutions}
\label{app:precess}

We want to make quantitative contact between our local plane-wave analysis and the global
motion of a freely precessing two-fluid neutron star. In particular, we need to
work out the two key quantities $\Omega_\rp^\parallel$ and $w_\parallel$,
which are required if we want to assess the relevance of the plane-wave instabilities.

The precession of two-fluid models, accounting for the mutual friction coupling,
has been discussed by \citet{swc}. As we will discuss in this Appendix,
it is
relatively straightforward to extract the  information we require from their
analysis.
They take as their starting point the two Euler equations
(\ref{eulern}) and (\ref{eulerp}), with the mutual friction force given by (\ref{mf}).
The precession equations are then obtained by assuming that each fluid
rotates as a solid body, with the two rotation axes misaligned.
The final equations of motion have a set of mode-solutions that represent
free precession damped by mutual friction. Here we will briefly summarise the
precession solutions in the weak and strong drag limits,
and then extract the quantities $\Omega_\rp^\parallel$ and $w_\parallel$
that enter into the plane-wave analysis. Having done this, we will consider an alternative approximation
which allows us to extend our analysis to (essentially) the entire range of
drag coefficients.

In the analysis of \citet{swc} it is assumed that the neutron star crust is
deformed in such a way that its moment of inertia tensor has diagonal components
$I_1=I_2=I_3/(1+\epsilon)$, where $\epsilon$ is small. Meanwhile, the superfluid
is assumed spherical and its moment of inertia is $I_s$. The calculation is then carried out in the crust frame,
considering perturbations around a background where $\Omega_\rn^i = \Omega_\rp^i = \Omega^i$.
Assuming that the precession modes have time dependence $\exp(p\Omega t)$ one finds that
$p$ must solve
\be
p^2 + \left[ i(1-\epsilon) + \tilde{I} ( {\cal B} -i {\cal B}^\prime ) \right] p 
+ \epsilon ( 1- {\cal B}' - i{\cal B}) = 0
\label{dispp}
\ee
where
\be
\tilde{I} = 1 + \frac{I_s}{I_1}
\ee
Once we have determined the precession period and the mutual friction damping timescale
from the roots of (\ref{dispp}), the eigenfunctions follow from 
equation~(74) in \citet{swc}. This provides the relation
\be
\bar{\Omega}_\rn = \left[ 1 - { p \over p + i(1-{\cal B}')+{\cal B}} \right] \bar{\Omega}_\rp
\label{Omrel}\ee
where we have defined 
\be
\bar{\Omega}_\rn = \Omega_{\rn,1}+i\Omega_{\rn,2} \ , \qquad
\bar{\Omega}_\rp = \Omega_{\rp,1}+i\Omega_{\rp,2}
\ee
and $\Omega_{\rn,1}$, for example, represents the $\hat{x}_1$ component of the angular velocity of the 
neutrons as measured in the crust frame. 


\subsection{The weak drag case}

In the weak drag case have ${\cal B} \ll 1$ and ${\cal B}' \approx
{\cal B}^2$.  As always, there are two mode solutions, one which
corresponds  to the familiar
Eulerian precession of a biaxial body in the limit ${\cal B} \rightarrow 0$, and the other to a less
familiar motion which has no analogue in one component systems.  We
will discuss each in turn, beginning with the Eulerian solution.  To
linear order in the mutual friction its frequency is given by 
\be p = i \epsilon - {\cal B} {\epsilon I_s \over (1+\epsilon) I_1}
\label{weaksol1}\ee
and the mode amplitudes in the proton and superfluid components are 
related by
\be
\bar{\Omega}_\rn = \left[ { 1 \over 1+\epsilon} 
- { i{\cal B} \epsilon \over (1+\epsilon)^3}
\left( 1 + \epsilon + {I_s \over I_1} \right) \right] \bar{\Omega}_\rp
\label{relate1}
\ee
Let us now assume that the eigenfunctions are normalized, i.e. 
choosing the precession amplitude and the initial phase, in such a way
that
\be
\bar{\Omega}_\rp = \alpha \Omega e^{p\Omega t}
\ee
with $\alpha$ real. Then we can immediately read off
\begin{eqnarray}
\Omega_{\rn,1} &=& \alpha \Omega e^{-t/\tau_d} \cos (2 \pi t/P_\mathrm{pr}) \\
\Omega_{\rn,2} &=& \alpha \Omega e^{-t/\tau_d} \sin (2\pi t/P_\mathrm{pr})
\end{eqnarray}
where the damping time is
\be
\tau_d = { 1 \over {\cal B} \epsilon \Omega} { I_1 \over I_s}
\ee
 while the precession period is
\be
P_\mathrm{pr} = 2 \pi / \epsilon \Omega
\ee
Using (\ref{relate1}) we find
\begin{multline}
\Omega_{\rn,1} = \alpha \Omega e^{-t/\tau_d} \Big[ { 1 \over 1+\epsilon} \cos (2 \pi t/P_\mathrm{pr})
\\
- { {\cal B} \epsilon \over (1+\epsilon)^3} \left( 1 + \epsilon + {I_s \over I_1} \right) \sin (2 \pi t/P_\mathrm{pr}) 
\Big]
\label{sol_weak1}
\end{multline}
\begin{multline}
\Omega_{\rn,2} = \alpha \Omega e^{-t/\tau_d} \Big[ { 1 \over 1+\epsilon} \sin (2 \pi t/P_\mathrm{pr})
\\
- { {\cal B} \epsilon \over (1+\epsilon)^3} \left( 1 + \epsilon + {I_s \over I_1} \right) \cos (2 \pi t/P_\mathrm{pr}) 
\Big]
\label{sol_weak2}
\end{multline}
It should also be recalled that we have $\Omega_{\rn,3}= \Omega_{\rp,3}= \Omega$.

We want to be able to use this solution as background for the
plane-wave calculation. For this to make sense we need the rotation to be
essentially constant. This is the case as long as we consider a timescale which is
short compared to both the damping timescale $\tau_d$ and the precession period $P_\mathrm{pr}$.
Since ${\cal B}$ is small, the strongest
constraint is provided by the precession timescale. Assume that $t \ll P_\mathrm{pr}$ we
can use
\be
\vec{\Omega}_\rp \approx \alpha \Omega \hat{x}_1 + \Omega \hat{x}_3
\ee
and
\be
\vec{\Omega}_\rn \approx { \alpha \Omega \over 1+ \epsilon} \hat{x}_1 + \Omega \hat{x}_3
\ee

One of the quantities we need for the
plane-wave analysis is $\Omega_\rp^\parallel$, which represents the projection of the
charged component rotation along the vortex array (along $\Omega_\rn^i$).
To work this out we need the norm of $\Omega_\rn^i$. We find that 
\begin{multline}
\hat{n} = { \vec{\Omega}_\rn \over \Omega_\rn } \approx 
\\
\approx { 1 \over (1 + \alpha^2)^{1/2} }
\left( 1 + {  \epsilon} { \alpha^2 \over 1 + \alpha^2} \right) \left[
{ \alpha \over 1+ \epsilon} \hat{x}_1 + \hat{x}_3 \right]
\end{multline}
This then leads to
\be
\Omega_\rp^\parallel \approx \left( 1 + {  \epsilon} { \alpha^2 \over 1 + \alpha^2} \right) \Omega_\rn
\ee
which shows that it is reasonable to take $\Omega_\rp^\parallel \approx \Omega_\rn$ in the plane-wave calculation.

We also need to estimate the relative linear velocity along the
vortex array, $w_\parallel$. We obtain this quantity from
\be
w_\parallel = (v_\rn^i - v_\rp^i) \hat{n}_i
\ee
where
\be
v_\rx^i = \epsilon^{ijk} \Omega^\rx_j x_k
\ee
In other words, the frame independent scalar quantity that we need
follows from
\be
w_\parallel = \epsilon^{ijk} \hat{n}_i ( \Omega^\rn_j - \Omega^\rp_j) x_k
\ee

From our mode solutions we find
\be
\vec{\Omega}_\rn - \vec{\Omega}_\rp \approx -  {\alpha \epsilon \Omega \over 1+\epsilon} \hat{x}_1
\ee
Then
\be
(\vec{\Omega}_\rn - \vec{\Omega}_\rp)\times \hat{n}
\approx { \epsilon \alpha \Omega \over (1+\alpha^2)^{1/2}} \left[ 1
+ {\epsilon \alpha^2 \over 1 + \alpha^2} \right] \hat{x}_2
\ee
and we see that
\be
w_\parallel \approx  {  \epsilon \alpha \Omega x_2}
\ee

Finally, we would like to be able to make direct contact with observations. This involves working out the 
``wobble angle'' $\theta_w$ associated with our precession solution. As in the analysis of \citet{dij}, we define 
the wobble angle to be the angle between the total angular momentum axis and the crust deformation axis. 
In the present case the latter is simply $\hat{x}_3$.
Meanwhile, in the crust frame, the total angular momentum is given by
\be
\vec{J} = (I_1 \Omega^\rp_1+ I_s \Omega^\rn_1) \hat{x}_1 + (I_1 \Omega^\rp_2 + I_s \Omega^\rn_2) \hat{x}_2
+ (I_3 + I_s) \Omega \hat{x}_3
\ee
Taking the modulus of this, we then find the wobble angle from
\be
\vec{J} \cdot \hat{x}_3 = (I_3 + I_s) \Omega = J \cos \theta_w
\ee
In the weak drag case we then have
\be
J \approx (I_s + I_3) \Omega \left[ 1  + {\alpha^2 \over 2 (1+\epsilon)^2} \right]
\ee
Hence
\be
\theta_w \approx { \alpha \over 1+ \epsilon} \approx \alpha
\label{wobble_weak}
\ee
This completes our analysis of the motion associated with (\ref{weaksol1}).


We now turn to the second mode found for weak mutual friction.  Its
frequency is given by
\begin{equation}
p = -i -  {\cal B} \frac{\tilde{I} + \epsilon}{1+\epsilon} ,
\label{weaksol2}\end{equation}
while the superfluid and proton mode amplitudes are related by
\begin{equation}
\bar{\Omega}_\rp =  i {\cal B} \frac{I_s}{I_3} \bar{\Omega}_\rn .
\end{equation}
To gain insight as to what sort of motion this mode corresponds to,
consider the case ${\cal B}= 0$.  Then we can write
\begin{eqnarray}
\bar{\Omega}_\rp &=& 0 ,\\
\bar{\Omega}_\rn &=& \alpha \Omega ,
\end{eqnarray}
i.e. the crust is unperturbed, and keeps spinning at a rate $\Omega$
about the inertial 3-axis, but the superfluid does something more
complicated.  Writing the superfluid components explicitly, we have
\begin{eqnarray}
\Omega_{\rn,1} &=& \alpha \Omega \cos(\Omega t) ,\\
\Omega_{\rn,2} &=& -\alpha \Omega \sin(\Omega t) .
\end{eqnarray}
It is then straightforward to show that the rate of change of
$\vec{\Omega}_\rn$ with respect to the \emph{inertial frame} is zero:
\begin{equation}
\frac{d \Omega_{\rn,i}}{dt_{\rm I}} = \frac{d \Omega_{\rn, i}}{dt_{\rm R}}
+ \epsilon_{ijk} \Omega^j \Omega_\rn^k = 0 ,
\end{equation}
where the subscripts `I' and `R' label time derivatives in the
inertial and rotating frames, respectively.  It follows that the
superfluid's angular velocity is fixed in space; in this limit the
mode corresponds to a misalignment between the crust and superfluid
spin axes by an angle $\alpha$.  The relative velocity between the
crust and superfluid is then
\begin{equation}
w_i = \epsilon_{ijk} \Omega_\rn^j x^k ,
\end{equation}
which gives a relative flow along the vortices at time $t=0$ of
\begin{equation}
w_\parallel = x_2 \alpha \Omega .
\end{equation}
This remains true to leading order in the case of small but non-zero
$\cal B$.

When ${\cal B} \neq 0$ we instead have
\begin{eqnarray}
\Omega_{\rn,1} &=&  \alpha \Omega \cos(\Omega t) e^{-t/\tau_d} , \\
\Omega_{\rn,2} &=& -\alpha \Omega  \sin(\Omega t) e^{-t/\tau_d} , \\
\Omega_{\rp,1} &=&  \alpha \Omega {\cal B} \frac{I_s}{I_3} \sin(\Omega t) e^{-t/\tau_d} 
 \ ,\\
\Omega_{\rp,2} &=& \alpha \Omega {\cal B} \frac{I_s}{I_3} \cos(\Omega t) e^{-t/\tau_d} 
\ ,
\end{eqnarray}
where the damping time $\tau_d$ is given by
\begin{equation}
\tau_d = \frac{1}{{\cal B}\Omega} \frac{1+\epsilon}{\tilde{I}+\epsilon} .
\end{equation}
The total angular momentum is the sum of the neutron and proton
contributions, which can be shown to be given by
\begin{eqnarray}
J_1 &=& \alpha \Omega e^{-t/\tau_d} I_s
\cos\left[\Omega t - \frac{\cal B}{1+\epsilon}\right] ,\\
J_2 &=& -\alpha \Omega e^{-t/\tau_d} I_s
\sin\left[\Omega t - \frac{\cal B}{1+\epsilon}\right] ,\\
J_3 &=& I_3 \Omega .
\end{eqnarray}
From these equations we can easily compute the angle between $J_i$ and
the crust's 3-axis: 
\begin{equation}
\cos^{-1}(\hat{J}\cdot \hat{x}_3) = \alpha e^{-t/\tau_d} \frac{I_s}{I_s+I_3} .
\end{equation}
The constancy of this angle (aside from its slow monotonic decrease)
shows that the crust's 3-axis moves on a cone about the fixed total
angular momentum.  To see exactly how the body moves it is best to
make use of Euler angles $(\theta,\phi,\psi)$ giving the orientation
of the body axes with respect to some fixed inertial axes.  Then the
angular velocity components of the crust referred to the rotating
crust frame take the form, see e.g. \citet{ll76},
\begin{eqnarray}
\Omega_{\rp,1} &=& \dot\phi \sin\theta \sin\psi + \dot\theta \cos\psi ,\\
\Omega_{\rp,2} &=& \dot\phi \sin\theta \cos\psi - \dot\theta \sin\psi ,\\
\Omega_{\rp,3} &=& \dot\phi \cos\theta + \dot\psi .
\end{eqnarray}
In defining $(\theta,\phi,\psi)$ we are free to make any choice of
fixed inertial frame we want, but obviously it would be simplest to
choose one where the inertial 3-axis of this system lies along the
fixed angular momentum vector.  Then $\theta = \cos^{-1}( \hat{J} \cdot
  \hat{x}_3)$ which is a constant, so the above equations reduce to
\begin{eqnarray}
\Omega_{\rp,1} &=& \dot\phi \sin\theta \sin\psi ,\\
\Omega_{\rp,2} &=& \dot\phi \sin\theta \cos\psi ,\\
\Omega_{\rp,3} &=& \dot\phi \cos\theta + \dot\psi ,
\end{eqnarray}
where
\begin{equation}
\theta = \alpha e^{-t/\tau_d} \frac{I_s}{I_s+I_3} .
\end{equation}
Now compare with
\begin{eqnarray}
\Omega_{\rp,1} &=&  A \sin\Omega t  ,\\
\Omega_{\rp,2} &=&  A \cos\Omega t  ,\\
\Omega_{\rp,3} &=& \Omega ,
\end{eqnarray}
where
\begin{equation}
A = \alpha \Omega e^{-t/\tau_d} {\cal B} { I_s \over I_3} .
\end{equation}
It follows at once that
\begin{equation}
A^2 = \dot\phi^2 \theta^2
\Rightarrow \dot\phi = \Omega {\cal B} \frac{I_s+I_3}{I_3} ,
\end{equation} 
and the $\Omega_{\rp,3}$ equation gives
\begin{equation}
\dot\psi = \Omega-\dot\phi = 
\Omega\left[1-{\cal B} \frac{I_s+I_3}{I_3}\right] .
\end{equation}
Collecting results:
\begin{eqnarray}
\theta &=& \alpha e^{-t/\tau_d} 
\frac{I_s}{I_s+I_3}  \hspace{5mm} \rm  \,\,\sim \, constant, \\
\dot\phi &=& {\cal B} \Omega \frac{I_s+I_3}{I_3} 
\hspace{20mm}\ll \Omega ,\\
\dot\psi &=& \Omega\left[1-{\cal B} \frac{I_s+I_3}{I_3}\right]
\hspace{8mm}\approx \Omega .
\end{eqnarray}
We therefore see that the motion is still rather simple, with the crust's
3-axis moving in a cone of half-angle $\theta$ about the fixed angular
momentum axis at the slow rate $\dot\phi \ll \Omega$, with a rapid
rotation at rate $\dot\psi \approx \Omega$ about the crust's 3-axis
superimposed.

Amusingly, this is the opposite of normal free precession, i.e.  small
angle free precession of a single component rigid biaxial (but nearly
spherical) body, as in that case $\dot\phi \approx \Omega$ and
$| \dot\psi | \approx \epsilon \Omega \ll \Omega$.  However, from the
point of view of the radio observations, both sorts of precession
consist of two superimposed rotations misaligned by a small angle and
so are observationally indistinguishable.  This means that rather than
interpreting PSR1828-11 as undergoing the usual precession mode with
the ratio $P/P_\mathrm{pr} \sim \epsilon$, we would instead have
\begin{equation}
\frac{P}{P_\mathrm{pr}} \approx \frac{\dot\phi}{\dot\psi} \approx
{\cal B} \frac{I_s+I_3}{I_3}
\end{equation} 
The potential physical significance of this is discussed in Section
\ref{sec:weak}.


\subsection{The strong drag case}

In the strong drag limit 
we have ${\cal B} \ll 1$ and $ 1-{\cal B}' \approx {\cal B}^2$. In this case we find the two precession modes
\be
p = - { \epsilon \over \bar{\sigma} } [ i (1-{\cal B}') + {\cal B}]
\label{strongsol1}\ee
a mode which damps out before it completes one full cycle, and
\be
p = i \bar{\sigma} - {\cal B} { 1 + \bar{\sigma} \over \bar{\sigma}} { I_s \over I_1}
\label{strongsol2}\ee
In these expressions we have used
\be
\bar{\sigma} = \epsilon + { I_s \over I_1}
\ee
The solution (\ref{strongsol2}) represents slowly damped fast precession, since $I_s/I_1 \sim 10 \gg \epsilon $. This is
essentially the precession mode discovered by \citet{shaham77} for a neutron star model where
an amount $I_s$ of neutron vortices is perfectly pinned.

Focusing on (\ref{strongsol2}) we find that
the damping time is 
\be
\tau_d = { \bar{\sigma} \over {\cal B} \Omega (1 + \bar{\sigma})} { I_1 \over I_s}
\ee
while the precession period is
\be
P_\mathrm{pr} = { 2 \pi \over \bar{\sigma} \Omega}
\ee
As in the weak coupling case the damping is relatively slow. The amplitudes are related by
\be
\bar{\Omega}_\rn = - i { {\cal B} \over \bar{\sigma} } \bar{\Omega}_\rp = - i {\cal B} \left( \epsilon
+ { I_s \over I_1} \right)^{-1} \bar{\Omega}_\rp
\ee
Again assuming that the plane-wave analysis is
relevant on a timescale short compared to $P_\mathrm{pr}$, we find the approximate mode solution
\be
\vec{\Omega}_\rp \approx \alpha \Omega \hat{x}_1 + \Omega \hat{x}_3
\label{sol_strong1}
\ee
\be
\vec{\Omega}_\rn \approx - \alpha {\cal B} \Omega
\left( \epsilon + { I_s \over I_1} \right)^{-1} \hat{x}_2 + \Omega \hat{x}_3
\label{sol_strong2}
\ee

In this case we see that
\be
\hat{n} \approx - \alpha {\cal B} \left( \epsilon + { I_s \over I_1} \right)^{-1}
\hat{x}_2 + \hat{x}_3
\ee
and it immediately follows that
\be
\Omega_\rp^\parallel \approx \Omega_\rn
\ee
as before.

To work out the relative flow along the vortex array, we take the same steps as in the weak drag case. This leads to
\be
\vec{\Omega}_\rn - \vec{\Omega}_\rp \approx - \alpha \Omega \hat{x}_1 - \alpha {\cal B} \Omega
\left( \epsilon + { I_s \over I_1} \right)^{-1} \hat{x}_2
\ee
and
\begin{multline}
(\vec{\Omega}_\rn - \vec{\Omega}_\rp)\times \hat{n}
\approx - \alpha {\cal B} \Omega  \left( \epsilon 
+ { I_s \over I_1} \right)^{-1} \hat{x}_1 \\
+ \alpha \Omega \hat{x}_2 -  \alpha^2 {\cal B} \Omega \left( \epsilon + { I_s \over I_1} \right)^{-1} \hat{x_3}
\end{multline}
which means that
\be
w_\parallel \approx  \alpha \Omega x_2
\ee

The final step again concerns the wobble angle.
For the above solution we find that
\be
J \approx (I_s+I_3) \Omega \left[ 1 + { \alpha^2 I_1^2 \over 2(I_s + I_3)^2} \right]
\ee
which leads to
\be
\theta_w \approx { \alpha I_1 \over I_s + I_3} \approx {\alpha I_1 \over I_s}
\label{wobble_strong}
\ee
since $\epsilon \ll I_s/I_1$


\subsection{The ``general'' case}

So far, the precession solutions that we have discussed are identical to the 
limiting cases considered by \citet{swc}. However, we can do 
better than this and find solutions that are valid across the
range of permissible drag coefficients, $\cal R$. The starting point is (\ref{dispp}) as before,
but now we assume that $\epsilon$ is the small parameter. 
This approximation is generally valid as long as $\mathcal{B}>>\epsilon$, i.e. it fails only in the extreme
weak drag limit (which we have already analyzed above). 

Considering (\ref{dispp}) to order $\epsilon$ we find the two solutions
\be
p_1 \approx  - \epsilon { 1-\mathcal{B}'-i\mathcal{B} \over  i (1-\tilde{I}\mathcal{B}') + \tilde{I}\mathcal{B}} \\
\label{gensol1}\ee
and
\be
p_2 \approx -  \tilde{I} \mathcal{B} - i \left( 1-\tilde{I} \mathcal{B}'\right)
+ \epsilon {I_s \over I_1} { i\mathcal{B}+\mathcal{B}' \over  i (1-\tilde{I}\mathcal{B}') + \tilde{I}\mathcal{B} }
\label{gensol2}
\ee
It is quite straightforward to show that these solution behave in the anticipated way in the 
weak- and strong drag limits. In particular, we find that (\ref{gensol1}) limits to
(\ref{weaksol1}) when $\mathcal{R}\to 0$ and to (\ref{strongsol1}) for $\mathcal{R}\to \infty$.
Meanwhile, (\ref{gensol2}) reduces to (\ref{weaksol2}) and to (\ref{strongsol2}) in the
respective limits.  

To extract the precession period and the damping time we need the real and imaginary parts of $p$.
The first solution (\ref{gensol1}) can be written
\be
p_1 \approx - { \epsilon \over \mathcal{D} } \left\{ - (1-\tilde{I})\mathcal{B}
-i \left[ \tilde{I}\mathcal{B}^2 + (1-\mathcal{B}')(1-\tilde{I}\mathcal{B}')\right] \right\}
\ee
where we have introduced 
\be
\mathcal{D} = \tilde{I}^2 \mathcal{B}^2 + (1 - \tilde{I} \mathcal{B}')^2
\ee
Since we assume that the perturbations behave as $e^{p\Omega t}$ we see that the damping time is
\be
\tau_d \approx { \mathcal{D} \over (\tilde{I}-1) \epsilon \mathcal{B} \Omega }
\ee
while the precession period is
\be
P_\mathrm{pr} \approx { 2\pi \over \epsilon\Omega}  { \mathcal{D} \over
 \tilde{I}\mathcal{B}^2 + (1-\mathcal{B}')(1-\tilde{I}\mathcal{B}')}
\ee

\begin{figure}
\centerline{\includegraphics[height= 7cm,clip]{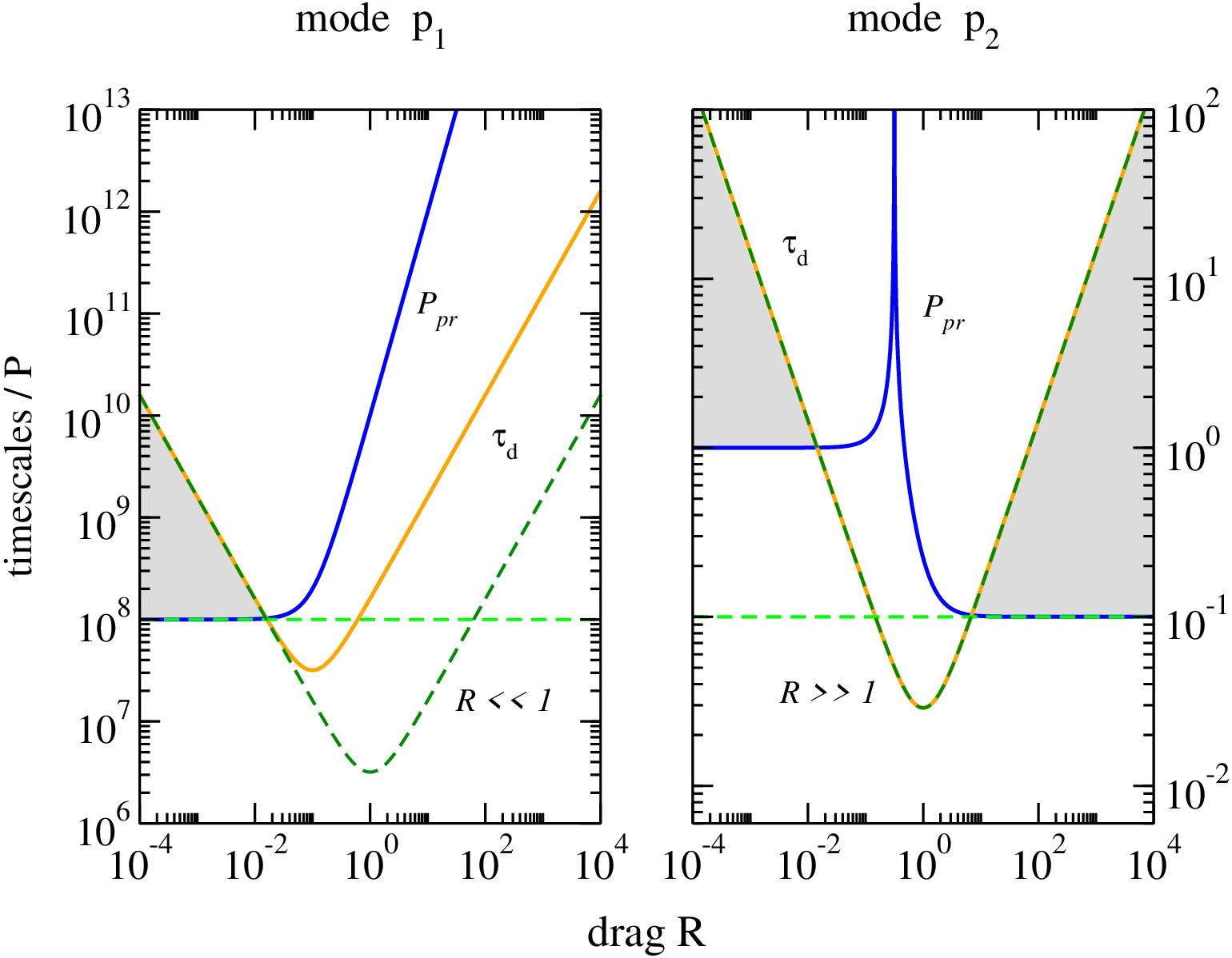}}
\caption{The solutions $\{P_{\rm pr}, \tau_d\}$ for the precession modes $p_1 $ (left panel) and $p_2$ (right panel)
for a general mutual friction drag ${\cal R}$. These are compared to the weak and strong  drag solutions, left and right
panel respectively, represented by the dashed curves. The shaded areas correspond to the intervals 
of ${\cal R} $ for which precession is not overdamped by mutual friction, i.e., $\tau_d > P_{\rm pr} $.
Note that all timescales are normalised with the rotational period $P$.} 
\label{overdamp}
\end{figure}

Given the above estimates we can determine the parameter range for which this precession mode is 
relevant. A reasonable criterion would be that the precession motion does not damp out before a
single cycle is completed. This means that we must have $\tau_d/P_\mathrm{pr}<1$. For the solution given above 
we find that this is true provided that 
\be
\mathcal{R} < { 1 \over 2 \pi} {I_1 \over I_s} \approx 0.016
\label{left}
\ee
for our canonical parameters, cf. the left panel of Figure~\ref{overdamp}. 
For larger values of the drag coefficient this solution is overdamped. 
Since the maximum value of $\mathcal R$ that we need to consider is small, one would expect the weak 
drag estimate to be sufficiently accurate for our analysis. As we can see in Figure~\ref{overdamp}
the weak drag solution matches the exact result well even before ${\cal R} $ enters the range (\ref{left}) .

Let us now move on to the second precession solution, (\ref{gensol2}). Here the algebra gets a little bit messier.
The frequency is given by
\be
p_2 \approx - \tilde{I} \mathcal{B} + \epsilon { I_s \mathcal{B} \over I_1\mathcal{D}}
 - i \left\{  (1- \tilde{I}\mathcal{B}') - \epsilon { I_s \over I_1 \mathcal{D} }
\left[ \tilde{I} \mathcal{B}^2 - (1-\tilde{I}\mathcal{B}') \mathcal{B}' \right] \right\}
\ee
from which we see that the precession period is
\bear
P_\mathrm{pr} \approx - { 2 \pi \over (1-\tilde{I}\mathcal{B}') \Omega } \left[ 1 + { \epsilon I_s \over I_1 \mathcal{D} }
{ \tilde{I}\mathcal{B}^2 - ( 1- \tilde{I}\mathcal{B}')\mathcal{B}' \over 1-\tilde{I}\mathcal{B}'} \right]
\eear
Meanwhile, the damping timescale is
\be
\tau_d \approx \frac{1}{\tilde{I}\Omega {\cal B}} \left ( 1 + \epsilon \frac{I_s}{I_1 \tilde{I}
{\cal D}} \right ) 
\ee

Again, we can assess the relevance of this mode by checking that it is not overdamped. The behaviour is illustrated 
in the right panel of Figure~\ref{overdamp}. In this case we find acceptable solution for 
\be
\mathcal{R} > 2\pi \left( 1 + {I_1 \over I_s} \right) \approx 7 
\ee
In this case $\mathcal R$ is sufficiently large (compared to unity) that one would expect 
the strong drag approximation to give reliable results. Indeed, it is clear from  Figure~\ref{overdamp}
that one would expect the strong drag approximation to be valid when this mode is
oscillatory.  In addition, the figure  shows that the mode is also relevant for ${\cal R} < 0.01 $. 
In that regime it is well approximated by the weak-drag solution (\ref{weaksol2}).

In practice,  the results for the general precession modes, $p_1$ and $p_2 $, shows that
the (much simpler) approximate weak and strong drag solutions  apply for the entire range 
of ${\cal R} $ for which precession is not overdamped.



\end{document}